\documentclass[12pt,a4paper]{article}

\usepackage{ifthen}
\newboolean{articletitles}
\setboolean{articletitles}{true} 
\newboolean{uprightparticles}
\setboolean{uprightparticles}{false} 
\newboolean{inbibliography}
\setboolean{inbibliography}{false} 

\usepackage{microtype}
\usepackage{lineno}  
\usepackage{xspace} 
\usepackage{caption} 

\usepackage{graphicx}  
\usepackage{color}
\usepackage{colortbl}
\graphicspath{{./figs/}} 

\usepackage{amsmath} 
\usepackage{amssymb}
\usepackage{amsfonts}
\usepackage{upgreek} 

\newcommand*\patchAmsMathEnvironmentForLineno[1]{%
\expandafter\let\csname old#1\expandafter\endcsname\csname #1\endcsname
\expandafter\let\csname oldend#1\expandafter\endcsname\csname
end#1\endcsname
 \renewenvironment{#1}%
   {\linenomath\csname old#1\endcsname}%
   {\csname oldend#1\endcsname\endlinenomath}%
}
\newcommand*\patchBothAmsMathEnvironmentsForLineno[1]{%
  \patchAmsMathEnvironmentForLineno{#1}%
  \patchAmsMathEnvironmentForLineno{#1*}%
}
\AtBeginDocument{%
\patchBothAmsMathEnvironmentsForLineno{equation}%
\patchBothAmsMathEnvironmentsForLineno{align}%
\patchBothAmsMathEnvironmentsForLineno{flalign}%
\patchBothAmsMathEnvironmentsForLineno{alignat}%
\patchBothAmsMathEnvironmentsForLineno{gather}%
\patchBothAmsMathEnvironmentsForLineno{multline}%
\patchBothAmsMathEnvironmentsForLineno{eqnarray}%
}

\usepackage{hyperref}    
\usepackage[all]{hypcap} 

\usepackage{xspace} 
\usepackage{upgreek}

\def\lhcb {\mbox{LHCb}\xspace}

\def\MagUp {\mbox{\em Mag\kern -0.05em Up}\xspace}

\ifthenelse{\boolean{uprightparticles}}%
{
 
 \def\Pgamma      {\ensuremath{\upgamma}\xspace}

 \def\Ppi         {\ensuremath{\uppi}\xspace}

 \def\PDelta      {\ensuremath{\Delta}\xspace}                 
 \def\PXi      {\ensuremath{\Xi}\xspace}                 
 \def\PLambda      {\ensuremath{\Lambda}\xspace}                 
 \def\PSigma      {\ensuremath{\Sigma}\xspace}                 
 \def\POmega      {\ensuremath{\Omega}\xspace}                 
 \def\PUpsilon      {\ensuremath{\Upsilon}\xspace}                 
 
 \def\PB      {\ensuremath{\mathrm{B}}\xspace}                 
                  
 \def\PD      {\ensuremath{\mathrm{D}}\xspace}

 \def\PK      {\ensuremath{\mathrm{K}}\xspace}

 \def\Pb      {\ensuremath{\mathrm{b}}\xspace}                 
 \def\Pc      {\ensuremath{\mathrm{c}}\xspace}

 \def\Pi      {\ensuremath{\mathrm{i}}\xspace}

 \def\Ps      {\ensuremath{\mathrm{s}}\xspace}

}
{
 
 \def\Pgamma      {\ensuremath{\gamma}\xspace}

 \def\Ppi         {\ensuremath{\pi}\xspace}

 \mathchardef\PDelta="7101
 \mathchardef\PXi="7104
 \mathchardef\PLambda="7103
 \mathchardef\PSigma="7106
 \mathchardef\POmega="710A
 \mathchardef\PUpsilon="7107
                  
 \def\PB      {\ensuremath{B}\xspace}                 
                  
 \def\PD      {\ensuremath{D}\xspace}

 \def\PK      {\ensuremath{K}\xspace}

 \def\Pb      {\ensuremath{b}\xspace}                 
 \def\Pc      {\ensuremath{c}\xspace}

 \def\Pi      {\ensuremath{i}\xspace}

 \def\Ps      {\ensuremath{s}\xspace}

}

\makeatletter
\ifcase \@ptsize \relax
  \newcommand{\miniscule}{\@setfontsize\miniscule{4}{5}}
\or
  \newcommand{\miniscule}{\@setfontsize\miniscule{5}{6}}
\or
  \newcommand{\miniscule}{\@setfontsize\miniscule{5}{6}}
\fi
\makeatother

\DeclareRobustCommand{\optbar}[1]{\shortstack{{\miniscule (\rule[.5ex]{1.25em}{.18mm})}
  \\ [-.7ex] $#1$}}

\def\g      {{\ensuremath{\Pgamma}}\xspace}

\def\squark    {{\ensuremath{\Ps}}\xspace}

\def\cquark    {{\ensuremath{\Pc}}\xspace}

\def\bquark    {{\ensuremath{\Pb}}\xspace}

\def\pion   {{\ensuremath{\Ppi}}\xspace}
\def\piz    {{\ensuremath{\pion^0}}\xspace}

\def\pip    {{\ensuremath{\pion^+}}\xspace}
\def\pim    {{\ensuremath{\pion^-}}\xspace}
\def\pipm   {{\ensuremath{\pion^\pm}}\xspace}
\def\pimp   {{\ensuremath{\pion^\mp}}\xspace}

\def\kaon    {{\ensuremath{\PK}}\xspace}
  \def\Kbar    {{\kern 0.2em\overline{\kern -0.2em \PK}{}}\xspace}

\def\KorKbar    {\kern 0.18em\optbar{\kern -0.18em K}{}\xspace}

\def\Kp      {{\ensuremath{\kaon^+}}\xspace}
\def\Km      {{\ensuremath{\kaon^-}}\xspace}
\def\Kpm     {{\ensuremath{\kaon^\pm}}\xspace}
\def\Kmp     {{\ensuremath{\kaon^\mp}}\xspace}
\def\KS      {{\ensuremath{\kaon^0_{\mathrm{ \scriptscriptstyle S}}}}\xspace}

  \def\Dbar    {{\kern 0.2em\overline{\kern -0.2em \PD}{}}\xspace}
\def\D       {{\ensuremath{\PD}}\xspace}

\def\DorDbar    {\kern 0.18em\optbar{\kern -0.18em D}{}\xspace}
\def\Dz      {{\ensuremath{\D^0}}\xspace}
\def\Dzb     {{\ensuremath{\Dbar{}^0}}\xspace}

\def\Dm      {{\ensuremath{\D^-}}\xspace}

\def\Dstarz  {{\ensuremath{\D^{*0}}}\xspace}

\def\Dstarp  {{\ensuremath{\D^{*+}}}\xspace}

\def\B       {{\ensuremath{\PB}}\xspace}
\def\Bbar    {{\ensuremath{\kern 0.18em\overline{\kern -0.18em \PB}{}}}\xspace}

\def\BorBbar    {\kern 0.18em\optbar{\kern -0.18em B}{}\xspace}
\def\Bz      {{\ensuremath{\B^0}}\xspace}

\def\Bu      {{\ensuremath{\B^+}}\xspace}
\def\Bub     {{\ensuremath{\B^-}}\xspace}
\def\Bp      {{\ensuremath{\Bu}}\xspace}
\def\Bm      {{\ensuremath{\Bub}}\xspace}
\def\Bpm     {{\ensuremath{\B^\pm}}\xspace}

\def\Bs      {{\ensuremath{\B^0_\squark}}\xspace}

  \def\Y#1S{\ensuremath{\PUpsilon{(#1S)}}\xspace}

\def\Lz          {{\ensuremath{\PLambda}}\xspace}
\def\Lbar        {{\ensuremath{\kern 0.1em\overline{\kern -0.1em\PLambda}}}\xspace}
\def\LorLbar    {\kern 0.18em\optbar{\kern -0.18em \PLambda}{}\xspace}

\def\Lb      {{\ensuremath{\Lz^0_\bquark}}\xspace}

\def\Lc      {{\ensuremath{\Lz^+_\cquark}}\xspace}

\def\to                 {\ensuremath{\rightarrow}\xspace}

\def\CP                {{\ensuremath{C\!P}}\xspace}

\def\AT#1     {\ensuremath{A_{\mathrm{T}}^{#1}}\xspace}           

\def\C#1      {\ensuremath{\mathcal{C}_{#1}}\xspace}                       
\def\Cp#1     {\ensuremath{\mathcal{C}_{#1}^{'}}\xspace}                    
\def\Ceff#1   {\ensuremath{\mathcal{C}_{#1}^{\mathrm{(eff)}}}\xspace}        
\def\Cpeff#1  {\ensuremath{\mathcal{C}_{#1}^{'\mathrm{(eff)}}}\xspace}       
\def\Ope#1    {\ensuremath{\mathcal{O}_{#1}}\xspace}                       
\def\Opep#1   {\ensuremath{\mathcal{O}_{#1}^{'}}\xspace}                    

\newcommand{\tev}{\ifthenelse{\boolean{inbibliography}}{\ensuremath{~T\kern -0.05em eV}\xspace}{\ensuremath{\mathrm{\,Te\kern -0.1em V}}}\xspace}
\newcommand{\gev}{\ensuremath{\mathrm{\,Ge\kern -0.1em V}}\xspace}
\newcommand{\mev}{\ensuremath{\mathrm{\,Me\kern -0.1em V}}\xspace}
\newcommand{\kev}{\ensuremath{\mathrm{\,ke\kern -0.1em V}}\xspace}
\newcommand{\ev}{\ensuremath{\mathrm{\,e\kern -0.1em V}}\xspace}
\newcommand{\gevc}{\ensuremath{{\mathrm{\,Ge\kern -0.1em V\!/}c}}\xspace}
\newcommand{\mevc}{\ensuremath{{\mathrm{\,Me\kern -0.1em V\!/}c}}\xspace}
\newcommand{\gevcc}{\ensuremath{{\mathrm{\,Ge\kern -0.1em V\!/}c^2}}\xspace}
\newcommand{\gevgevcccc}{\ensuremath{{\mathrm{\,Ge\kern -0.1em V^2\!/}c^4}}\xspace}
\newcommand{\mevcc}{\ensuremath{{\mathrm{\,Me\kern -0.1em V\!/}c^2}}\xspace}

\def\mum  {\ensuremath{{\,\upmu\mathrm{m}}}\xspace}

\def\gsim{{~\raise.15em\hbox{$>$}\kern-.85em
          \lower.35em\hbox{$\sim$}~}\xspace}
\def\lsim{{~\raise.15em\hbox{$<$}\kern-.85em
          \lower.35em\hbox{$\sim$}~}\xspace}

\def\ptot       {\mbox{$p$}\xspace}
\def\pt         {\mbox{$p_{\mathrm{ T}}$}\xspace}

\def\evtgen     {\mbox{\textsc{EvtGen}}\xspace}

\def\geant      {\mbox{\textsc{Geant4}}\xspace}

\def\photos     {\mbox{\textsc{Photos}}\xspace}

\def\pythia     {\mbox{\textsc{Pythia}}\xspace}

\def\tell1  {TELL1\xspace}
\def\ukl1   {UKL1\xspace}

\usepackage{cite} 
\usepackage{mciteplus}

\usepackage{placeins}

\begin{document}

\renewcommand{\thefootnote}{\fnsymbol{footnote}}
\setcounter{footnote}{1}

\begin{titlepage}
\pagenumbering{roman}

\vspace*{-1.5cm}
\centerline{\large EUROPEAN ORGANIZATION FOR NUCLEAR RESEARCH (CERN)}
\vspace*{1.5cm}
\noindent
\begin{tabular*}{\linewidth}{lc@{\extracolsep{\fill}}r@{\extracolsep{0pt}}}
\vspace*{-1.2cm}\mbox{\!\!\!\includegraphics[width=.12\textwidth]{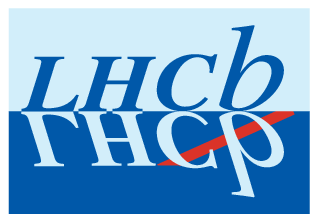}} & &
\\
 & & CERN-EP-2016-065 \\  
 & & LHCb-PAPER-2016-003 \\  
 & & \today \\ 
 & & \\
\end{tabular*}

\vspace*{2.0cm}

{\normalfont\bfseries\boldmath\huge
\begin{center}
Measurement of $C\!P$ observables in $B^\pm \to D K^\pm$ and $B^\pm \to D \pi^\pm$ with two- and four-body $D$ decays
\end{center}
}

\vspace*{1.0cm}

\begin{center}
The LHCb collaboration\footnote{Authors are listed at the end of this Letter.}
\end{center}

\vspace{\fill}

\begin{abstract}
\noindent
Measurements of $C\!P$ observables in $B^\pm \rightarrow D K^\pm$ and $B^\pm \rightarrow D \pi^\pm$ decays are presented where the $D$ meson is reconstructed in the final states $K^\pm\pi^\mp$, $\pi^\pm K^\mp$, $K^+K^-$, $\pi^+\pi^-$, $K^\pm\pi^\mp \pi^+ \pi^-$, $\pi^\pm K^\mp \pi^+ \pi^-$ and $\pi^+ \pi^- \pi^+ \pi^-$. 
This analysis uses a sample of charged $B$ mesons from $pp$ collisions collected by the LHCb experiment in 2011 and 2012, corresponding to an integrated luminosity of 3.0 fb$^{-1}$. 
Various \CP-violating effects are reported and together these measurements provide important input for the determination of the unitarity triangle angle $\gamma$. 
The analysis of the four-pion $D$ decay mode is the first of its kind.
\end{abstract}

\vspace*{1.0cm}

\begin{center}
  Published in Phys.~Lett.~B~760 (2016) Pages 117-131
\end{center}

\vspace{\fill}

{\footnotesize 
\centerline{\copyright~CERN on behalf of the \lhcb collaboration, licence \href{http://creativecommons.org/licenses/by/4.0/}{CC-BY-4.0}.}}
\vspace*{2mm}

\end{titlepage}

\newpage
\setcounter{page}{2}
\mbox{~}

\cleardoublepage

\renewcommand{\thefootnote}{\arabic{footnote}}
\setcounter{footnote}{0}

\pagestyle{plain}
\setcounter{page}{1}
\pagenumbering{arabic}

\section{Introduction}
\label{sec:Introduction}
 
A set of overconstraining measurements of the unitarity triangle from the CKM matrix is central to the validation of the Standard Model (SM) description of \CP violation~\cite{Cabibbo:1963yz,*Kobayashi:1973fv}.
Of these, the least-well measured is the angle $\gamma \equiv \text{arg}(-V^{}_{ud}V_{ub}^{*}/V^{}_{cd}V_{cb}^{*})$ with a precision, from a combination of measurements, of about $7^\circ$; this may be compared with the $3^\circ$ and $<1^\circ$~precision on the other angles $\alpha$ and $\beta$~\cite{Charles:2004jd,Bona:2006ah}. 
Amongst the three angles, \g is unique in that it does not depend on a coupling to the top quark and thus may be studied at tree level, largely avoiding possible influence from non-SM \CP violation.

The most powerful method for determining $\gamma$ in tree-level decays is through measurement of relative partial widths in $\Bm \to D\Km$ decays, where $D$ represents a \Dz or \Dzb meson.\footnote{The inclusion of charge-conjugate processes is implied except in any discussion of asymmetry.} 
The amplitude for the $\Bm \to \Dz \Km$ contribution is proportional to $V_{cb}$ while the amplitude for $\Bm \to \Dzb \Km$ is proportional to $V_{ub}$. 
By reconstructing hadronic \D decays accessible to both \Dz and \Dzb mesons, phase information may be extracted from the interference of the two amplitudes. 
The size of the resulting direct \CP violation is governed by the magnitude of the ratio $r_B$ of the $b\to u\bar{c}s$ amplitude to the $b\to c\bar{u}s$ amplitude.
The relatively large value of $r_B$ (about 0.1) in $\Bm \to D\Km$ decays means that the relative phase of the two interfering amplitudes can be obtained. 
This relative phase has a \CP-violating (weak) contribution and \CP-conserving (strong) contribution $\delta_B$; 
a measurement of the total phase for both \Bp and \Bm disentangles \g and $\delta_B$. 
Similar interference effects occur in $\Bpm \to D\pipm$ decays, albeit with reduced sensitivity to the phases because, due to additional Cabibbo suppression factors, the ratio of amplitudes is about 20 times smaller.

The study of $\Bm \to D\Km$ decays for measurements of $\gamma$ was first suggested for \CP eigenstates of the \D decay, for example the \CP-even $\D \to\Kp\Km$ and $\D \to\pip\pim$ decays, labelled here GLW modes~\cite{Gronau:1990ra,Gronau1991172}. 
The argument has been extended to suppressed $\D \to\pim\Kp$ decays where the interplay between the favoured and suppressed decay paths in both the \Bm and the neutral \D decays results in a large charge asymmetry.
This is the so-called ADS mode~\cite{Atwood:1996ci}, which introduces a dependency on the ratio of the suppressed and favoured \D decay amplitudes $r_D$ and their  phase difference $\delta_D$.
The $\Bm \to [h^+h^-]_D h^-$ ADS/GLW decays $(h = K,\pi)$ have been studied at the \B factories~\cite{Lees:2013nha,PhysRevLett.106.231803} and at LHCb~\cite{Aaij:1432548}. This letter contains the updated and improved result using both the 2011 and 2012 data samples. 
The 2012 data benefits from a higher \Bpm meson production cross-section and a more efficient trigger, so this update is approximately a factor four increase in statistics.

The  ADS/GLW formalism can be extended to four-particle \D decays.
However, there are multiple intermediate resonances with differing amplitude ratios and strong phases with the consequence that the interference in the \Bm decay, and hence the sensitivity to \g, is diluted~\cite{Atwood:2003mj}. 
For $\D \to\Km\pip\pip\pim$ and $\D \to\pim\Kp\pip\pim$ decays this dilution is parameterised in terms of a coherence factor $\kappa^{K3\pi}$, an effective strong phase difference averaged over all contributing resonances $\delta_D^{K3\pi}$, and an overall suppressed-to-favoured amplitude ratio $r_D^{K3\pi}$. 
Best sensitivity to \g is achieved using independent measurements of the $\kappa^{K3\pi}$ and $\delta_D^{K3\pi}$ parameters, which have been determined using a sample of quantum-correlated $\Dz\Dzb$ pairs~\cite{Evans:2016tlp,Lowery:2009id}, and by the study of \D-mixing in this final state~\cite{Aaij:2016rhq}.
A similar dilution parameter, labelled the \CP fraction $F_+^{4\pi}$ can be defined for $\D \to \pip\pim\pip\pim$ decays~\cite{Nayak:2014tea}. 
For this final state it is found that  $F_+^{4\pi}=0.737\pm0.028$~\cite{Malde:2015mha}, so that the decay behaves like a \CP-even GLW mode, albeit with the interference effects reduced by a factor $(2F_+^{4\pi}-1)\approx0.5$. 

This letter includes an analysis of \mbox{$\Bm \to [h^+h^-\pip\pim]_D h^-$} decays and supersedes the previous analysis of \mbox{$\Bm \to [\pim\Kp\pip\pim]_D h^-$~\cite{Aaij:1529913}} and complements the study of the \mbox{$\Bm \to [h^+h^-\piz]_D h^-$} modes~\cite{Aaij:2015jna}. 
The analysis of the four-pion $D$ decay mode is the first of its kind.
In total, 21 measurements of \CP observables are reported. Two of these are ratios of the favoured $\Bm \to \Dz\Km$ and $\Bm \to \Dz\pim$ partial widths,
\begin{equation}
R_{K/\pi}^{f} = \frac{\Gamma(\Bm \to [f]_{D}\Km) + \Gamma(\Bp \to [\bar f]_{\D}\Kp)}{\Gamma(\Bm \to [f]_{D}\pim) + \Gamma(\Bp \to [\bar f]_{\D}\pip)}\,,
\end{equation}
where $f$ is $\Km \pip (\pim \pip)$ and $\bar{f}$ is its charge-conjugate state. Three are double ratios that are sensitive to the partial widths of the (quasi-)GLW modes, $f=\pip\pim(\pip\pim)$ and $\Kp\Km$, normalised to those of the favoured modes of the same multiplicity,
\begin{equation}
R^{KK} = \frac{R_{K/\pi}^{KK}}{R_{K/\pi}^{K\pi}}\,, \ \ \ \ \ 
R^{\pi\pi} = \frac{R_{K/\pi}^{\pi\pi}}{R_{K/\pi}^{K\pi}}\,, \ \ \ \ \ 
R^{\pi\pi\pi\pi} = \frac{R_{K/\pi}^{\pi\pi\pi\pi}}{R_{K/\pi}^{K\pi\pi\pi}}\,.
\end{equation}
Five observables are charge asymmetries,
\begin{equation}
A_{h}^{f} = \frac{\Gamma(\Bm \to [f]_{D}h^{-}) - \Gamma(\Bp \to [\bar f]_{\D}h^{+})}{\Gamma(\Bm \to [f]_{D}h^{-}) + \Gamma(\Bp \to [\bar f]_{\D}h^{+})}\,,
\end{equation}
for $h=K$ and $f=\Km \pip (\pim \pip),~\pip\pim(\pip\pim)$ and $\Kp\Km$. There are a further three asymmetries for $h=\pi$ and $f=\pip\pim(\pip\pim)$ and $\Kp\Km$.
Four observables are partial widths of the suppressed ADS modes relative to their corresponding favoured decays,
\begin{equation}
R_{\text{ADS}(h)}^{\bar f} = \frac{\Gamma(\Bm \to [\bar f]_{D} h^-) + \Gamma(\Bp \to [f]_{\D} h^+)}{\Gamma(\Bm \to [f]_{D} h^-) + \Gamma(\Bp \to [\bar f]_{\D} h^+ )}\,,
\end{equation}
with which come four ADS-mode charge asymmetries,
\begin{equation}
A_{\text{ADS}(h)}^{\bar f} = \frac{\Gamma(\Bm \to [\bar f]_{D}h^{-}) - \Gamma(\Bp \to [f]_{D}h^{+})}{\Gamma(\Bm \to [\bar f]_{D}h^{-}) + \Gamma(\Bp \to [f]_{D}h^{+})}\,.
\end{equation}
An alternative formulation of the ADS observables measures the suppressed ADS modes relative to their favoured counterparts, independently for \Bp and \Bm mesons,
\begin{equation}
R_{+(h)}^{\bar f} = \frac{\Gamma(\Bp \to [f]_{\D} h^+)}{\Gamma(\Bp \to [\bar f]_{\D} h^+ )}\,, \ \ \ \ \ \ 
R_{-(h)}^{\bar f} = \frac{\Gamma(\Bm \to [\bar f]_{D} h^-)}{\Gamma(\Bm \to [f]_{D} h^-)}\,.
\end{equation}
All the charge asymmetry measurements are affected by a possible asymmetry in the \Bpm production cross-section multiplied by any overall asymmetry  from the LHCb detector, together denoted as $\sigma^\prime$. 
This effective production asymmetry, defined as $A_{\Bpm} = \textstyle{\frac{\sigma^\prime(\Bm)-\sigma^\prime(\Bp)}{\sigma^\prime(\Bm)+\sigma^\prime(\Bp)}}$, is measured in this analysis from the charge asymmetry of the most abundant $\Bm \to [\Km \pip]_{D}\pim$ and $\Bm \to [\Km \pim\pip\pim]_{D}\pim$ modes. 
This measurement is applied as a correction to all other \CP asymmetry results. 
In these modes, the possible \CP asymmetry, as derived from existing knowledge of \g and $r_B$ in this decay~\cite{Aaij:2013zfa}, is smaller than the uncertainty on existing measurements of the \Bpm production asymmetry~\cite{Aaij:2012jw}. 
The \CP asymmetry is thus assumed to be zero with a small systematic uncertainty.
Remaining detection asymmetries, notably between \Km and \Kp, are corrected for using calibration samples.

\section{Detector and simulation}
\label{sec:Detector}

The \lhcb detector~\cite{Alves:2008zz,*LHCb-DP-2014-002} is a single-arm forward
spectrometer covering the \mbox{pseudorapidity} range $2<\eta <5$,
designed for the study of particles containing \bquark or \cquark
quarks. The detector includes a high-precision tracking system
consisting of a silicon-strip vertex detector surrounding the $pp$
interaction region, a large-area silicon-strip detector located
upstream of a dipole magnet with a bending power of about
$4{\mathrm{\,Tm}}$, and three stations of silicon-strip detectors and straw
drift tubes placed downstream of the magnet.
The tracking system provides a measurement of momentum, \ptot, of charged particles with
a relative uncertainty that varies from 0.5\% at low momentum to 1.0\% at 200\gevc.
The minimum distance of a track to a primary vertex (PV), the impact parameter (IP), is measured with a resolution of $(15+29/\pt)\mum$,
where \pt is the component of the momentum transverse to the beam, in\,\gevc.
Different types of charged hadrons are distinguished using information
from two ring-imaging Cherenkov detectors. 
Photons, electrons and hadrons are identified by a calorimeter system consisting of
scintillating-pad and preshower detectors, an electromagnetic
calorimeter and a hadronic calorimeter. Muons are identified by a
system composed of alternating layers of iron and multiwire
proportional chambers.
The trigger consists of a hardware stage, based on information from the calorimeter and muon
systems, followed by a software stage, in which all charged particles
with $\pt>500\,(300)\mev$ are reconstructed for 2011\,(2012) data.

  At the hardware trigger stage, events are required to have a muon with high \pt or a
  hadron, photon or electron with high transverse energy in the calorimeters. For hadrons,
  the transverse energy threshold is 3.5\gev.
  The software trigger requires a two-, three- or four-track
  secondary vertex with significant displacement from the primary
  $pp$ interaction vertices. At least one charged particle
  must have transverse momentum $\pt > 1.7\gevc$ and be
  inconsistent with originating from a PV.
  A multivariate algorithm~\cite{BBDT} is used for
  the identification of secondary vertices consistent with the decay
  of a \bquark hadron.

In the simulation, $pp$ collisions are generated using
\pythia\,8~\cite{Sjostrand:2007gs,*Sjostrand:2006za} 
 with a specific \lhcb
configuration~\cite{LHCb-PROC-2010-056}.  Decays of hadronic particles
are described by \evtgen~\cite{Lange:2001uf}, in which final-state
radiation is generated using \photos~\cite{Golonka:2005pn}. The
interaction of the generated particles with the detector, and its response,
are implemented using the \geant
toolkit~\cite{Allison:2006ve, *Agostinelli:2002hh} as described in
Ref.~\cite{LHCb-PROC-2011-006}.

\section{Event selection}
\label{sec:Selection}

After reconstruction of the \D meson candidate from either two or four charged particles, the same basic event selection is applied to all $\Bm \to D h^-$ channels of interest. 
The reconstructed \D meson candidate mass is required to be within $\pm 25 \mevcc$ of its known value~\cite{PDG2014}. 
This mass range corresponds to approximately three times the mass resolution of the signal peaks. 
The kaon or pion originating from the \Bpm decay, subsequently referred to as the bachelor particle, is required to have \pt in the range \mbox{$0.5 - 10.0$ \gevc} and $p$ in the range \mbox{$5 - 100$ \gevc}. 
These requirements ensure that the track is within the kinematic coverage of the RICH detectors, which are used to provide particle identification (PID) information. Details of the PID calibration procedure are given in Sect.~\ref{sec:Fit}. 
In addition, a kinematic fit is performed to each decay chain, with vertex constraints applied to both the \Bpm and \D vertices, and the \D candidate constrained to its known mass~\cite{Hulsbergen:2005pu}. 
Events are required to have been triggered by either the decay products of the signal candidate or particles produced elsewhere in the $pp$ collision.
The \Bpm meson candidates with an invariant mass in the interval \mbox{$5079 - 5899$ \mevcc} are retained. 
Each \Bpm candidate is associated to the PV to which it has the smallest IP. 

For both the two- and four-body \D-mode selections, a pair of boosted decision tree (BDT) discriminators, implementing the gradient boost algorithm~\cite{Roe}, are employed to achieve further background suppression. 
The BDTs are trained using simulated \mbox{$\Bm \to [\Km \pip (\pip\pim)]_{D}\Km$} decays together with a background sample of $K\pipm$ combinations with invariant mass in the range \mbox{$5900-7200$ \mevcc}. 
For the first BDT, those backgrounds with a \D candidate mass more than $\pm 30$ \mevcc away from the known \Dz mass are used in the training. In the second BDT, backgrounds with a \D candidate mass within $\pm25$ \mevcc of the known \Dz mass are used. 
A loose cut on the classifier response of the first BDT is applied prior to training the second one. This focusses the second BDT training on backgrounds enriched with fully reconstructed \D mesons. 

The input to the BDT is a set of quantities that characterise the signal decay. These quantities can be divided into two categories:
(1) properties of any particle and (2) properties of composite particles only (the \D and \Bpm candidates). Specifically:
\begin{enumerate}
\item{$p$, \pt and the square of the IP significance;}
\item{decay time, flight distance, decay vertex quality, radial distance between the decay vertex and the PV, and the angle between the particle's momentum vector and the line connecting the production and decay vertex.}
\end{enumerate} 
Signal purity is improved by using a variable that estimates the imbalance of \pt around the \Bpm candidate, defined as
\begin{equation}
I_{\pt} = \frac{\pt(\Bpm) - \Sigma \pt}{\pt(\Bpm) + \Sigma \pt}\,,
\end{equation}
where the sum is taken over tracks lying within a cone around the \Bpm candidate, excluding the tracks related to the signal. 
The cone is defined by a circle with a radius of 1.5 units in the plane of pseudorapidity and azimuthal angle (expressed in radians). 
The BDT thus gives preference to \Bpm candidates that are either isolated from the rest of the event, or consistent with a recoil against another $b$ hadron. 

No PID information is used in the BDT training so the efficiency for $\Bm \to D\Km$ and $\Bm \to D \pim$ decays is similar, with insignificant variations arising from the small differences in the kinematics. 
The cuts on the two BDT selections are optimised by minimising the expected uncertainty on $A_{\text{ADS}(K)}^{\pi K (\pi\pi)}$, as measured in the invariant mass fit described below.
The purity of the sample is further improved with RICH information by requiring all kaons and pions in the \D decay to be correctly identified with a PID selection that has an efficiency of about 85\% per particle.



Peaking backgrounds from charmless decays are suppressed by requiring that the flight distance significance of the \D candidate from the \Bpm decay vertex is larger than two standard deviations. 
The residual charmless contribution is interpolated from fits to the \Bpm mass spectrum (without the kinematic fit of the decay chain) in both the lower and upper \D-mass sidebands. 
The charmless yields are determined independently for \Bp and \Bm candidates and are later used in the mass fit as fixed terms, with their uncertainties included in the systematic uncertainties of the final results. 
The largest residual charmless contributions are in the $\Bm \to [\pip \pim (\pip \pim)]_{D}\Km$ modes which show charge-integrated yields of $88 \pm 11$ and $115 \pm 11$ for the two- and four-pion modes. This is 7\% and 8\% of the measured signal yields.

Even with PID requirements, the suppressed ADS samples contain significant cross-feed from favoured signal decays where the \Km and a \pip from the \D decay are misidentified as a \pim and \Kp. 
This contamination is reduced by removing any candidate whose reconstructed \D mass, under the exchange of mass hypotheses between the kaon and an opposite-sign pion, lies within $\pm15 \mevcc$ of the known \Dz mass. This veto is also applied to the favoured mode, with the same efficiency. The residual cross-feed rates are estimated in data from the favoured sample, assuming the veto and PID efficiencies factorise;
they are $(4.3 \pm 0.2)\times 10^{-5}$ for $\Bm \to [\Km \pip]_{D}h^-$ and $(2.7 \pm 0.1)\times 10^{-4}$  for $\Bm \to [\Km \pip \pip \pim]_{D}h^-$.

After the above selections, multiple candidates exist in 0.1\% and 1\% of events in the \mbox{$\Bm \to [h^+h^-]_{D}h^-$} and \mbox{$\Bm \to [h^+h^-\pip\pim]_{D}h^-$} samples, respectively.
Only one candidate per event is retained for the main fit. When more than one candidate is selected, the one with the best \Bpm vertex quality is retained.


\section{Signal yields and systematic uncertainties}
\label{sec:Fit}

The values of the \CP observables are determined using binned maximum-likelihood fits to the invariant mass distributions of selected \Bpm candidates. 
Independent fits are used for the $\Bm \to [h^+h^-]_{D} h^-$ and $\Bm \to [h^+h^-\pip\pim]_{D} h^-$ samples. 
Information from the RICH detectors is used to separate $\Bm \to D \Km$ from $\Bm \to D \pim$ decays with a PID requirement on the bachelor particle. 
Distinguishing between \Bp and \Bm candidates, bachelor particle hypotheses, and four (three) \D daughter final states, yields 16 (12) disjoint samples in the \mbox{$\Bm \to [h^+h^-]_{D} h^-$} (\mbox{$\Bm \to [h^+h^-\pip\pim]_{D} h^-$}) fit, which are fitted simultaneously. 
The total probability density function (PDF) is built from two signal PDFs, for $\Bm \to D \Km$ and $\Bm \to D \pim$ decays, and three types of background PDF. 
All PDFs are identical for \Bp and \Bm decays.

\begin{enumerate}
\item{
{\boldmath$\Bm \to D\pim$}

In the $D\pim$ samples an asymmetric double-Gaussian-like function is used for the $\Bpm \to D\pipm$ signal,
\begin{equation}
f(m) = f_{\text{core}}\text{ exp}\left(\frac{-(m-\mu)^{2}}{2\sigma_c^{2} + (m-\mu)^{2}\alpha_{L,R}}\right) + (1-f_{\text{core}})\text{ exp}\left(\frac{-(m-\mu)^{2}}{2\sigma_{w}^{2}}\right)\,, \label{eq:Cruijff}
\end{equation}
which has a peak position $\mu$ and core width $\sigma_c$, where $\alpha_{L}(m < \mu)$ and \mbox{$\alpha_{R}(m > \mu)$} parameterise the tails. 
The $\mu$ and $\alpha$ parameters are shared across all samples but the core width parameter varies independently for each \D final state, except in the suppressed $\pi K(\pi\pi)$ PDFs which are required to be identical to their favoured $K\pi(\pi\pi)$ counterpart. 
The additional Gaussian function with a small fractional contribution of about 1\% is found necessary to model satisfactorily the tails of the peak. 

The $\Bm \to D\pim$ decays misidentified as $\Bm \to D\Km$ are displaced to higher mass in the $D\Km$ subsamples. 
These misidentified candidates are modelled by the sum of two Gaussian functions with common mean but modified to include tail components as in Eq.~\ref{eq:Cruijff}. 
The mean, widths and one tail parameter are left to vary freely. 
} 

\item{
{\boldmath$\Bm \to D\Km$}

In the $D\Km$ samples, Eq.~\ref{eq:Cruijff} is again used for the signal PDF. 
The peak position $\mu$ and the two tail parameters $\alpha_{L}$ and $\alpha_{R}$ are fixed to those of the $\Bm \to D \pim$ signal function, as are the wide component parameters $f_{\text{core}}$ and $\sigma_{w}$. 
The core width parameter in each \D mode is related to the corresponding $\Bm \to D \pim$ width by a freely varying ratio common to all \D final states. 
Misidentified $\Bpm \to D\Kpm$ candidates appear in the $D\pim$ subsamples and are described by a fixed shape obtained from simulation, which is later varied to determine a systematic uncertainty associated with this choice.

} 

\item{
{\textbf{Combinatorial background}}

Due to the low background level, a linear function is sufficient to describe the entire invariant mass spectrum. 
Two common slope parameters are used, one for $D\pim$ and another for $D\Km$ subsamples but yields vary independently.
} 

\item{
{\textbf{Peaking backgrounds}}

Charmless \Bpm decay and the favoured mode cross-feed backgrounds both peak at the \Bpm mass and are indistinguishable from the signal.
Their residual yields are estimated in data, entering the fit as fixed proportions of the favoured $\Bm \to D \pim$ yield. 
A Gaussian function is used for the PDF, with a $(25\pm2)~\mevcc$ width parameter that is taken from simulation; this is about 50\% wider than the signal PDF.
}

\item{
{\textbf{Partially reconstructed \boldmath{$b$}\,-hadron decays}}

Partially reconstructed backgrounds generally have lower invariant mass than the signal peak. 
The dominant contributions are from $\Bm \to D h^- \piz$, $\Bm \to \Dstarz \pim$ and $\Bz \to \Dstarp \pim$ decays where either a photon or a pion is missed in the reconstruction.
The distribution of each of these sources in the invariant mass spectrum depends on the spin and mass of the missing particle. 
If the missing particle has spin-parity $0^{-}\ (1^{-})$, the distribution is parameterised by a parabola with positive (negative) curvature convolved with a Gaussian resolution function.
The kinematics of the decay that produced the missing particle define the endpoints of the range of the parabola.
Decays in which both a particle is missed and a bachelor pion is misidentified as a kaon are parameterised with a semi-empirical PDF, formed from the sum of Gaussian and error functions. The parameters of each partially reconstructed PDF are fixed to the values found in fits to simulated events, and are varied as a source of systematic uncertainty. The yields of each contribution vary independently in each subsample, where all partially reconstructed decay modes share a common effective charge asymmetry across all \D modes. Though its effect is mitigated by the limited range of the mass fit, large \CP violation in the low-mass background is possible in the GLW and ADS samples, so a systematic uncertainty is assigned.

In the $\Bm \to [\Kp \Km]_{D} h^-$ samples, $\Lb \to [p^+ \Km \pip]_{\Lc} h^-$ decays contribute to background when the pion is missed and the proton is misidentified as the second kaon. 
The wide PDF of this component is fixed from simulation but the yield in the $\Bm \to [\Kp \Km]_{D} \pim$ subsample varies freely. 
The $\Lb \to [p^+ \Km \pip]_{\Lc} \Km$ yield is constrained using a recent measurement of $\mathcal{B}(\Lb\to\Lc\Km)/\mathcal{B}(\Lb\to\Lc\pim)$~\cite{Aaij:2013pka}. Furthermore, \mbox{$\Bs \to \Dzb \Km \pip$} decays, where the pion is missed, form a background for the suppressed $\Bm \to \D \Km$ modes. 
The yield of this component varies in the fit but the PDF is taken from a simulation model of the three-body \Bs decay~\cite{Aaij:2014baa}, smeared to match the resolution measured in data. 

} 

\end{enumerate}

In the $D\Km$ subsamples, the $\Bpm \to D \pim$ cross-feed can be determined by the fit to data. 
The $\Bm \to D\Km$ cross-feed into the $D\pim$ subsamples is not well separated from background,
so the expected yield is determined by a PID calibration procedure using approximately 20 million $\Dstarp \to [\Km \pip]_{D} \pip$ decays. 
The clean reconstruction of this charm decay is performed using kinematic variables only and thus provides a high purity sample of \Kmp and \pipm tracks, unbiased in the PID variables. 
The PID efficiency depends on track momentum and pseudorapidity, as well as the number of tracks in the event. 
The effective PID efficiency of the signal is determined by weighting the calibration sample such that the distributions of these variables match those of the selected candidates in the $\Bm \to D \pim$ mass distribution. 
It is found that 68.0\% ($\epsilon_{\rm PID(K)}$) of $\Bm \to D\Km$ decays pass the bachelor kaon PID requirement; the remaining 32.0\% cross-feed into the $\Bm \to D \pim$ sample.
With this selection, approximately 98\% of the $\Bm \to D \pim$ decays are correctly identified.
Due to the size of the calibration sample, the statistical uncertainty is negligible; the systematic uncertainty of the method is determined by the size of the signal track samples used, and thus increases for the lower statistics modes. The systematic uncertainty on $\epsilon_{\rm PID(K)}$ ranges from 0.3\% in $\Bm \to [\Km\pip]_{D}\Km$ to 1.5\% in $\Bm \to [\pip\pim\pip\pim]_{D}\Km$. 

\begin{figure}[ht]
  \begin{center}
    \includegraphics*[width=0.9\textwidth]{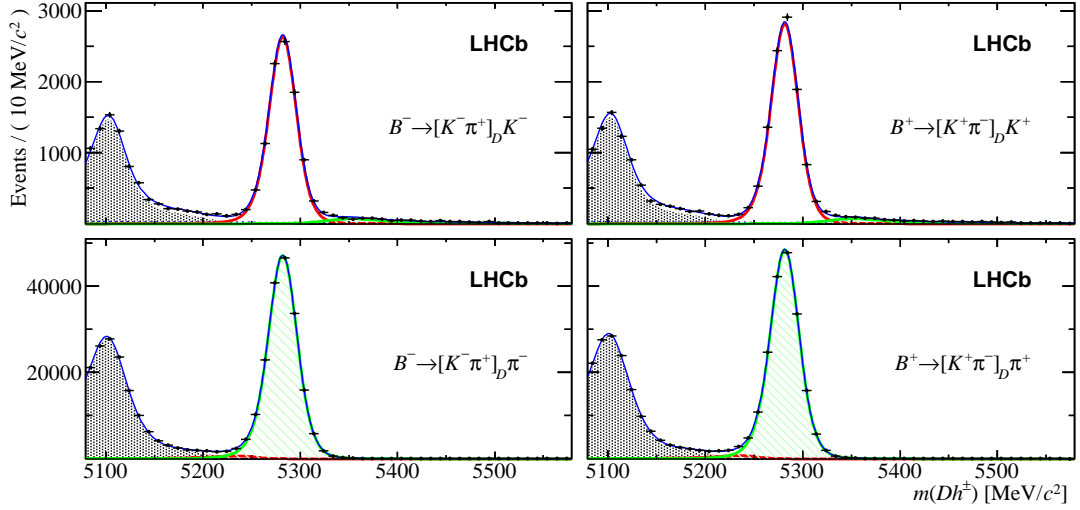}
  \caption{Invariant mass distributions of selected $\Bpm \to [\Kpm \pimp]_{D}h^{\pm}$ candidates, separated
by charge, with $\Bm{\rm(}\Bp{\rm)}$ candidates on the left\,(right). 
The top plots contain the $\Bpm \to DK^{\pm}$ candidate sample, as defined by a PID requirement on the bachelor particle.
The remaining candidates are placed in the bottom row, reconstructed with a pion hypothesis for the bachelor. 
The red (thick, open) and green (hatched-area) curves represent the $\Bpm \to D \Kpm$ and $\Bpm \to D \pipm$ signals. 
The shaded part indicates partially reconstructed decays, the dotted line, where visible, shows the combinatorial component, and the total PDF is drawn as a thin blue line.
  \label{fig:fit_kpi}}
  \end{center}
\end{figure}

\begin{figure}[ht]
  \begin{center}
    \includegraphics*[width=0.9\textwidth]{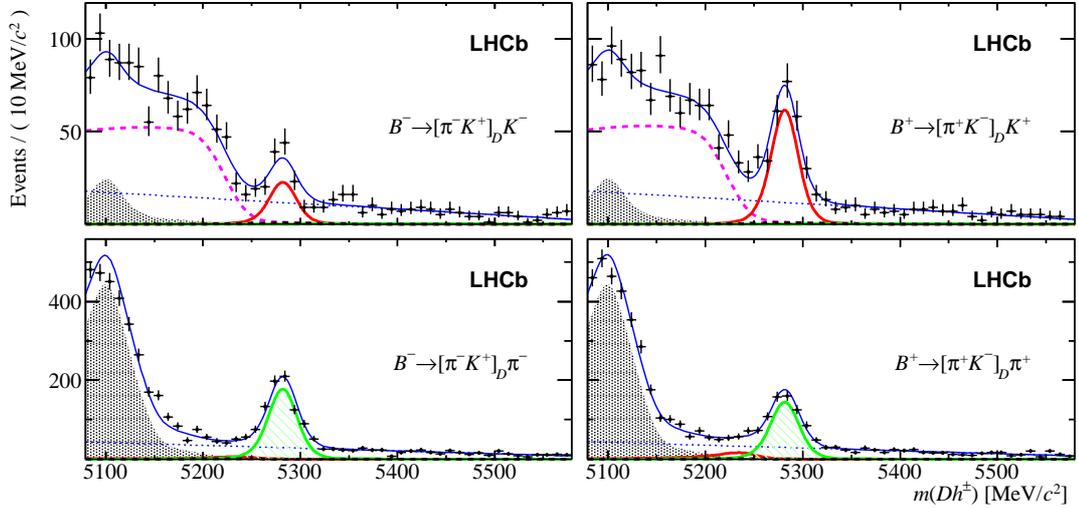}
  \caption{Invariant mass distributions of selected $\Bpm \to [\pipm \Kmp]_{D}h^{\pm}$ decays, separated by charge. 
The dashed pink line left of the signal peak shows partially reconstructed \mbox{$\B_{s}^{0} \to [\Kp\pim]_D \Km \pip$} decays, where the bachelor pion is missed. 
The favoured mode cross-feed is also included in the fit, but is too small to be seen.
See the caption of Fig.~\ref{fig:fit_kpi} for other definitions.
  \label{fig:fit_pik}}
  \end{center}
\end{figure}

\begin{figure}[ht]
  \begin{center}
    \includegraphics*[width=0.9\textwidth]{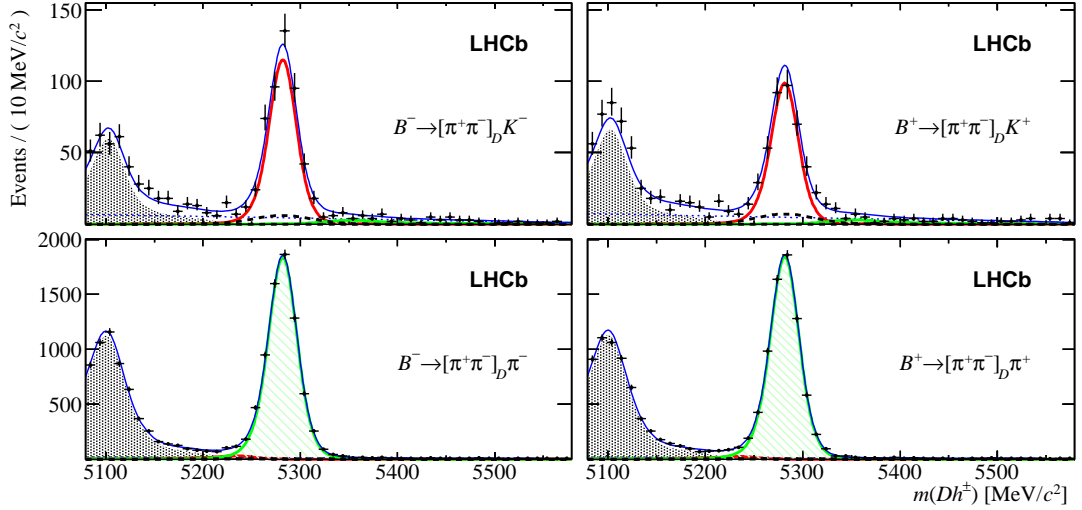}
  \caption{Invariant mass distributions of selected $\Bpm \to [\pip \pim]_{D}h^{\pm}$ candidates, separated by charge. 
The dashed black line represents the residual contribution from charmless decays. 
This component is present in the $D$ final states considered, but is most visible in this case.
See the caption of Fig.~\ref{fig:fit_kpi} for other definitions.
  \label{fig:fit_pipi}}
  \end{center}
\end{figure}

\begin{figure}[ht]
  \begin{center}
    \includegraphics*[width=0.9\textwidth]{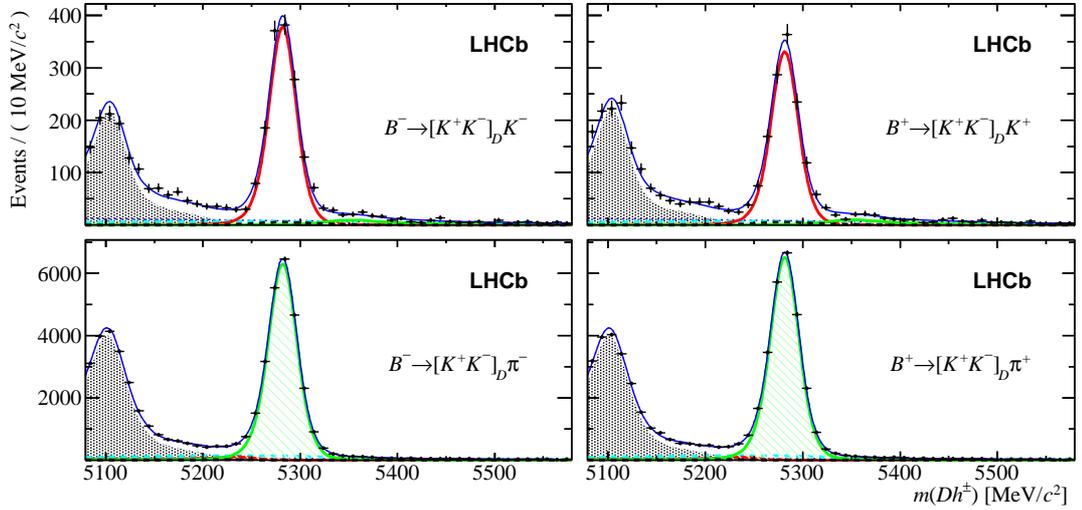}
  \caption{Invariant mass distributions of selected $\Bpm \to [\Kp \Km]_{D}h^{\pm}$ candidates, separated by charge. 
The dashed cyan line represents partially reconstructed $\Lb \to [p^+ \Km \pip]_\Lc h^-$ decays, where the pion is missed and the proton is misidentified as a kaon.
See the caption of Fig.~\ref{fig:fit_kpi} for other definitions.
  \label{fig:fit_kk}}
  \end{center}
\end{figure}

\begin{figure}[ht]
  \begin{center}
    \includegraphics*[width=0.9\textwidth]{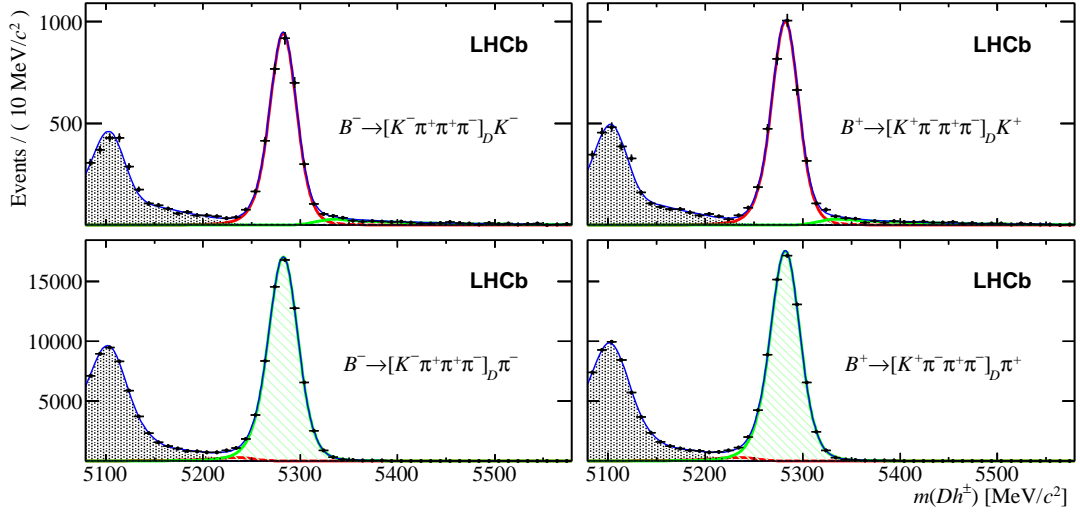}
  \caption{Invariant mass distributions of selected $\Bpm \to [\Kpm \pimp \pip \pim]_{D}h^{\pm}$ candidates, separated by charge. 
  See the caption of Fig.~\ref{fig:fit_kpi} for the definitions.
  \label{fig:fit_kpipipi}}
  \end{center}
\end{figure}

\begin{figure}[ht]
  \begin{center}
    \includegraphics*[width=0.9\textwidth]{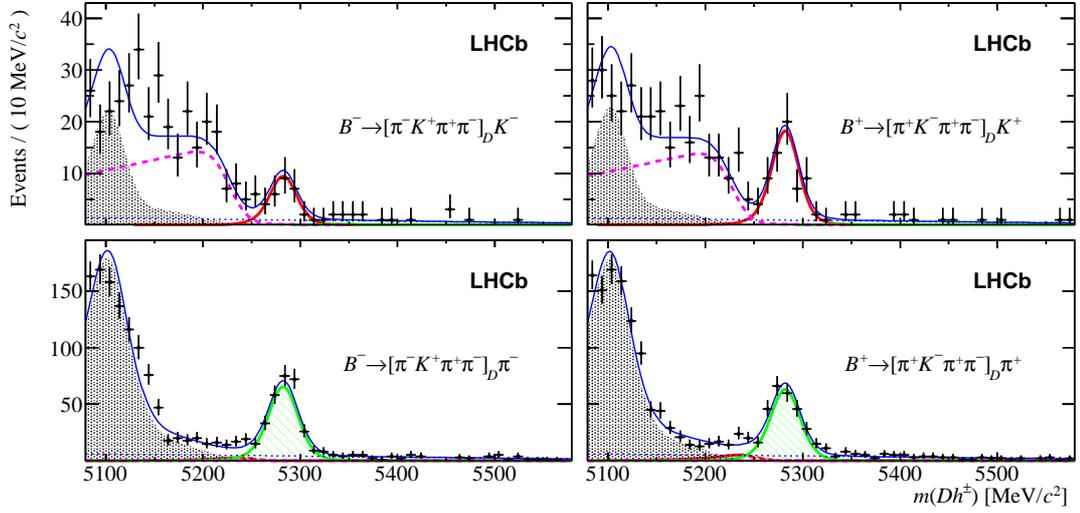}
  \caption{Invariant mass distributions of selected $\Bpm \to [\pipm \Kmp \pip \pim]_{D}h^{\pm}$ candidates, separated by charge. 
  The dashed pink line left of the signal peak shows partially reconstructed \mbox{$\B_{s}^{0} \to [\Kp\pim\pip\pim]_D \Km \pip$} decays, where the bachelor pion is missed. 
See the caption of Fig.~\ref{fig:fit_kpi} for other definitions.
\label{fig:fit_pikpipi}}
  \end{center}
\end{figure}

\begin{figure}[ht]
  \begin{center}
    \includegraphics*[width=0.9\textwidth]{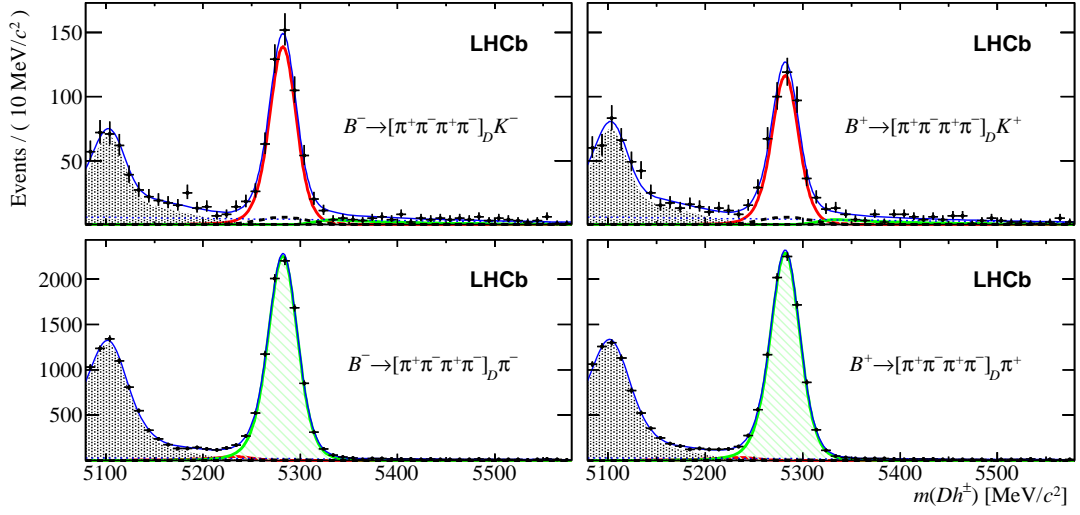}
  \caption{Invariant mass distributions of selected $\Bpm \to [\pip \pim \pip \pim]_{D}h^{\pm}$ candidates, separated by charge. 
The dashed black line represents the residual contribution from charmless decays. 
See the caption of Fig.~\ref{fig:fit_kpi} for other definitions.
\label{fig:fit_pipipipi}}
  \end{center}
\end{figure}

In order to measure \CP asymmetries, the detection asymmetries for \Kpm and \pipm must be taken into account. 
A detection asymmetry of $(-0.96 \pm 0.10)$\% is assigned for each kaon in the final state, arising from the fact that the nuclear interaction length of \Km mesons is shorter than that of \Kp mesons. 
This is computed by comparing the charge asymmetries in $\Dm\to\Kp\pim\pim$ and $\Dm\to\KS\pim$ calibration samples and weighting to the kinematics of the signal kaons. 
The equivalent asymmetry for pions is smaller $(-0.17 \pm 0.10)$\% and is taken from Ref.~\cite{LHCb-PAPER-2014-013}. 
The \CP asymmetries in the favoured $\Bm \to [\Km \pip (\pip\pim)]_{D}\pim$ decays are fixed to zero, with a systematic uncertainty of 0.16\% calculated from existing knowledge of \g and $r_B$ in this decay~\cite{Aaij:2013zfa}, with no assumption made about the strong phase. 
This enables the effective production asymmetry, $A_{\Bpm}$, to be measured and simultaneously subtracted from the charge asymmetry measurements in other modes. 
The signal yield for each mode is a sum of the number of signal and cross-feed candidates; their values are given in Table~\ref{yields}. 
The corresponding invariant mass spectra, separated by charge, are shown in \mbox{Figs.~\ref{fig:fit_kpi}$-$\ref{fig:fit_pipipipi}.}

To obtain the observables $R_{K/\pi}^{f}$, the ratio of yields must be corrected by the relative efficiency with which $\Bm \to D\Km$ and $\Bm \to D\pim$ decays are reconstructed and selected. 
From simulation, this ratio is found to be $1.017 \pm 0.017$ and $1.018 \pm 0.026$ for the two- and four-body \D decay selections. 
The uncertainties are calculated from the finite size of the simulated samples and account for imperfect modelling of the relative pion and kaon absorption in the detector material.
\begin{table}[b]
   \centering
      \small
    \caption{Signal yields as measured in the $\Bm \to [h^+h^-]_{D}h^-$ and $\Bm \to [h^+h^-\pip\pim]_{D}h^-$ invariant mass fits, together with their statistical uncertainties. \label{yields} }
    \begin{tabular}{l r l }
         Decay mode & \multicolumn{2}{c}{Yield} \\ \hline
{$\Bpm \rightarrow \left[ \Kpm \pimp \right]_{\D} \pipm$} & 378,050\!\!\!\!&$\pm$\phantom{0}650 \\
{$\Bpm \rightarrow \left[ \Kpm \pimp \right]_{\D} \Kpm$} & \phantom{0}29,470\!\!\!\!&$\pm$\phantom{0}230 \\
{$\Bpm \rightarrow \left[ \Kp \Km \right]_{\D} \pipm$} & \phantom{0}50,140\!\!\!\!&$\pm$\phantom{0}270 \\
{$\Bpm \rightarrow \left[ \Kp \Km \right]_{\D} \Kpm$} & \phantom{00}3816\!\!\!\!&$\pm$\phantom{0}92  \\
{$\Bpm \rightarrow \left[ \pip \pim \right]_{\D} \pipm$} & \phantom{0}14,680\!\!\!\!&$\pm$\phantom{0}130 \\
{$\Bpm \rightarrow \left[ \pip \pim \right]_{\D} \Kpm$} & \phantom{00}1162\!\!\!\!&$\pm$\phantom{0}48 \\
{$\Bpm \rightarrow \left[ \pipm K^{\mp} \right]_{\D} \pipm$} & \phantom{00}1360\!\!\!\!&$\pm$\phantom{0}44 \\
{$\Bpm \rightarrow \left[ \pipm K^{\mp} \right]_{\D} \Kpm$} & \phantom{000}553\!\!\!\!&$\pm$\phantom{0}34 \\
{$\Bpm\to\left[\Kpm\pimp\pip\pim\right]_D\pipm$}& 142,910\!\!\!\!&$\pm$\phantom{0}390\\
{$\Bpm\to\left[\Kpm\pimp\pip\pim\right]_D\Kpm$}& \phantom{0}11,330\!\!\!\!&$\pm$\phantom{0}140\\
{$\Bpm\to\left[\pip\pim\pip\pim\right]_D\pipm$}& \phantom{0}19,360\!\!\!\!&$\pm$\phantom{0}150\\
{$\Bpm\to\left[\pip\pim\pip\pim\right]_D\Kpm$}& \phantom{00}1497\!\!\!\!&$\pm$\phantom{0}60\\
{$\Bpm\to\left[\pipm\Kmp\pip\pim\right]_D\pipm$}& \phantom{000}539\!\!\!\!&$\pm$\phantom{0}26\\
{$\Bpm\to\left[\pipm\Kmp\pip\pim\right]_D\Kpm$}& \phantom{000}159\!\!\!\!&$\pm$\phantom{0}17
     \end{tabular}
\end{table}

The 21 observables of interest are free parameters of the fit.
The systematic uncertainties associated with fixed external parameters are assessed by repeating the fit many times, varying the value of each external parameter according to a Gaussian distribution within its uncertainty. The resulting spread (RMS) in each observable's value is taken as the systematic uncertainty on that observable due to the external source. 
The systematic uncertainties, grouped into four categories, are listed in Tables~\ref{systematics_2body} and~\ref{systematics_4body} for the two-body and four-body \D mode fits. 
Correlations between the categories are negligible and the total systematic uncertainties are given by the sums in quadrature. 

\begin{table}[h]
\renewcommand{\tabcolsep}{5.5pt}
   \begin{center}
   \scriptsize
    \caption{Systematic uncertainties for the $\Bm \to [h^+h^-]_{D}h^-$ \CP observables quoted as a percentage of the statistical uncertainty on the observable. PID refers to the PID calibration procedure. Bkg refers to the choice of background shapes and yields in the fit. Sim refers to the use of finite samples of simulated events to determine efficiency ratios. Asym refers to the fixed pion and kaon detection asymmetries, and the assumption of no \CP violation in $\Bm \to \Dz\pim$ decays.  \label{systematics_2body} }
    \begin{tabular}{l | c c c c c c c c c c c c c}
         [\%] & $A_{K}^{K\pi}$ & $R_{K/\pi}^{K\pi}$ & $A_{K}^{KK}$ & $A_{\pi}^{KK}$ & $R^{KK}$ & $A_{K}^{\pi\pi}$ & $A_{\pi}^{\pi\pi}$ & $R^{\pi\pi}$ & $R_{\text{ADS}(\pi)}^{\pi K}$ & $R_{\text{ADS}(K)}^{\pi K}$ & $A_{\text{ADS}(\pi)}^{\pi K}$ & $A_{\text{ADS}(K)}^{\pi K}$   \\ \hline
         PID & 42& 95 & 11 & 1     & 38 & 9 & 9 & 39 & 29 & 25 & 15 & 5 \\
         Bkg & 65& 190 & 34 & 3   & 84 & 30 & 28 & 48 & 69 & 74 & 24 & 15 \\       
         Sim & 21& 250 & 14 & 0   & 24 & 8 & 7 & 13 & 29 & 30 & 8 & 5 \\     
         Asym & 23& 27 & 11 & 34 & 6 & 7 & 20 & 5 & 12 & 13 & 7 & 8 \\ \hline
         Total & 83 & 330 & 40 & 34 & 96 & 33 & 36 & 64 & 81 & 85 & 30 & 19    
         \end{tabular}
\vspace{10mm}
    \caption{Systematic uncertainties for the $\Bm \to [h^+h^-\pip\pim]_{D}h^-$ \CP observables quoted as a percentage of the statistical uncertainty on the observable. See the Table~\ref{systematics_2body} caption for definitions.  \label{systematics_4body}}
    \vspace{-3mm}
     \begin{tabular}{l | c c c c c c c c c c}
     [\%] & $R_{K/\pi}^{K\pi\pi\pi}$ & $R^{\pi\pi\pi\pi}$ & $R_{\text{ADS}(K)}^{\pi K\pi\pi}$ & $R_{\text{ADS}(\pi)}^{\pi K\pi\pi}$ & $A_{K}^{K\pi\pi\pi}$ & $A_{\text{ADS}(K)}^{\pi K\pi\pi}$ & $A_{\text{ADS}(\pi)}^{\pi K\pi\pi}$ & $A_{K}^{\pi\pi\pi\pi}$ & $A_{\pi}^{\pi\pi\pi\pi}$ \\ \hline
     PID & 37 & 43 & 1 & 2 & 1 & 1 & 0 & 0 & 1 \\
     Bkg & 63 & 28 & 40 & 33 & 2 & 36 & 8 & 54 & 21 \\
     Sim & 160 & 0 & 0 & 0 & 0 & 1 & 0 & 0 & 0 \\
     Asym & 20	 & 5 & 7 & 6 & 16 & 5 & 5 & 8 & 22 \\ \hline
     Total & 180 & 51 & 41 & 34 & 16 & 36 & 10 & 54 & 30 \\
 \end{tabular}
 \end{center}
\end{table}

\FloatBarrier

\section{Results}
\label{sec:results}

The results of the fits to data, with statistical and systematic uncertainties, are:
\begin{align*}
A_{K}^{K\pi}  &= -0.0194\phantom{0}  \pm 0.0072\phantom{0}  \pm 0.0060\phantom{0}    \\ 
R_{K/\pi}^{K\pi} &= \phantom{-}0.0779\phantom{0}   \pm 0.0006\phantom{0}   \pm 0.0019\phantom{0}   \\
A_{K}^{KK} &= \phantom{-}0.087\phantom{00}  \pm 0.020\phantom{00}  \pm 0.008\phantom{0}   \\ 
A_{\pi}^{KK}  &= -0.0145\phantom{0}  \pm 0.0050\phantom{0}  \pm 0.0017\phantom{0}    \\ 
R^{KK} &= \phantom{-}0.968\phantom{00}  \pm 0.022\phantom{00}  \pm 0.021\phantom{00}  \\
A_{K}^{\pi\pi} &= \phantom{-}0.128\phantom{00}  \pm 0.037\phantom{00}  \pm 0.012\phantom{00}   \\ 
A_{\pi}^{\pi\pi} &= \phantom{-} 0.0043 \phantom{0}\pm 0.0086\phantom{0} \pm 0.0031  \\
R^{\pi\pi} &= \phantom{-}1.002\phantom{00} \pm 0.040\phantom{00} \pm 0.026\phantom{00}  \\ 
R_{\text{ADS}(\pi)}^{\pi K} &= \phantom{-}0.00360 \pm 0.00012 \pm 0.00009  \\
R_{\text{ADS}(K)}^{\pi K} &= \phantom{-}0.0188\phantom{0}  \pm 0.0011\phantom{0}  \pm 0.0010\phantom{0}   \\
A_{\text{ADS}(\pi)}^{\pi K} &= \phantom{-}0.100\phantom{00}  \pm 0.031\phantom{00}  \pm 0.009\phantom{0}   \\
A_{\text{ADS}(K)}^{\pi K} &= -0.403\phantom{00}  \pm 0.056\phantom{00}  \pm 0.011\,,
\end{align*}
\begin{align*}
R_{K/\pi}^{K\pi\pi\pi} &= \phantom{-}0.0793\phantom{0}  \pm 0.0010\phantom{0}  \pm 0.0018\phantom{0}   \\
R^{\pi\pi\pi\pi} &= \phantom{-}0.975\phantom{00}  \pm 0.037\phantom{00}  \pm 0.019\phantom{00}   \\
R_{\text{ADS}(K)}^{\pi K\pi\pi} &= \phantom{-}0.0140\phantom{0} \pm 0.0015\phantom{0} \pm 0.0006  \\
R_{\text{ADS}(\pi)}^{\pi K\pi\pi} &= \phantom{-}0.00377 \pm 0.00018\pm 0.00006  \\
A_{K}^{K\pi\pi\pi} &= \phantom{-}0.000\phantom{00}  \pm 0.012\phantom{00}  \pm 0.002\phantom{0}   \\
A_{\text{ADS}(K)}^{\pi K\pi\pi} &= -0.313\phantom{00}  \pm 0.102\phantom{00}  \pm 0.038\phantom{00}   \\
A_{\text{ADS}(\pi)}^{\pi K\pi\pi} &= \phantom{-}0.023\phantom{00}  \pm 0.048\phantom{00}  \pm 0.005\phantom{0}   \\
A_{K}^{\pi\pi\pi\pi} &= \phantom{-}0.100\phantom{00}  \pm 0.034\phantom{00}  \pm 0.018\phantom{0}   \\
A_{\pi}^{\pi\pi\pi\pi} &= -0.0041\phantom{0} \pm 0.0079\phantom{0} \pm 0.0024\,.
\end{align*}
These results supersede those in Refs. \cite{Aaij:1432548} and \cite{Aaij:2015jna} except for the results relating to the four-pion \D decay, which are reported for the first time.
The correlation matrices are given in the Appendix. 
The statistical correlations for the observables $R_{\text{ADS}}$ and $A_{\text{ADS}}$ are small. 
The alternative ADS observables are calculated: 
\mbox{$R_{+(K)}^{\pi K} = (2.58\pm0.23)\%$;}
\mbox{$R_{-(K)}^{\pi K} = (1.15\pm0.14)\%$;}
\mbox{$R_{+(\pi)}^{\pi K} = (3.22\pm0.18)\times10^{-3}$;}
\mbox{$R_{-(\pi)}^{\pi K} = (3.98\pm0.19)\times10^{-3}$;}
\mbox{$R_{+(K)}^{\pi K\pi\pi} = (1.82\pm0.25)\%$;}
\mbox{$R_{-(K)}^{\pi K\pi\pi} = (0.98\pm0.20)\%$;}
\mbox{$R_{+(\pi)}^{\pi K\pi\pi} = (3.68\pm0.26)\times10^{-3}$;} and
\mbox{$R_{-(\pi)}^{\pi K\pi\pi} = (3.87\pm0.26)\times10^{-3}$},
where the 	quoted uncertainties combine statistical and systematic effects.

The asymmetries in the two \CP-even \D decays, $\D\to\Kp\Km$ and $\D\to\pip\pim$, are averaged by noting that their systematic uncertainties are nearly fully correlated,
\begin{align*}
A_{\CP(K)}&= 0.097\pm0.018\pm0.009\,, \ \ A_{\CP(\pi)}= -0.0098\pm0.0043\pm0.0021\,.
\end{align*}
Similarly the average ratio of partial widths from these \D modes is
\begin{align*}
R^{\langle KK,\pi\pi\rangle}&= 0.978\pm0.019\pm0.018\,(\pm\,0.010)\,;
\end{align*}
in this case the systematic uncertainties, which are dominated by different background estimations, are only weakly correlated.
The third uncertainty arises only when the simplifying assumption is made that $r_{B}=0$ in $\Bm\to\D\pim$ decays. 
In this case, $R^{\langle KK,\pi\pi\rangle}$ becomes equal to the classic GLW observable $R_{\CP(K)}$~\cite{Gronau1991172}.
This additional uncertainty is applicable to the $KK$ and $\pi\pi$ modes individually but the equivalent uncertainty for the four-pion mode is lower, $\pm0.005$, due to the reduced coherence in that \D decay.
 
The significance of these measurements may be quantified from the likelihood ratio with respect to a \CP-symmetric null hypothesis, $\sqrt{-2\log(\mathcal{L}_0/\mathcal{L})}$. 
The significance of \CP violation in the ADS mode $\Bm \to [\pim\Kp]_D\Km$ is $8.0\,\sigma$ (standard deviations) and represents the first observation of \CP violation in a single $\Bm\to\D h^-$ decay mode.  
The combination of the two GLW modes \mbox{$\Bm \to [\Kp\Km]_D\Km$} and \mbox{$\Bm \to [\pip\pim]_D\Km$} also demonstrates a \CP-violation effect with $5.0\,\sigma$ significance. 
Taken together, the ADS and GLW modes of the \mbox{$\Bm \to D\pim$} decays show evidence of \CP violation with $3.9\,\sigma$ significance after accounting for systematic uncertainties. The $\Bm \to [\pip\pim\pip\pim]_D\Km$ data shows a $2.7\,\sigma$ \CP-violation effect.

 \section{Acceptance effects}
 \label{sec:acc}
The non-uniform acceptance across the phase space of the four-body modes affects the applicability of the external coherence factor and strong phase difference measurements~\cite{Evans:2016tlp} in the interpretation of these results.
With an acceptance model for the four-body \D decays from simulation, the effective values of the $D\to \Kp\pim\pip\pim$ coherence parameters are calculated using a range of plausible amplitude models. 
The acceptance is found to be almost uniform and the effective values of the coherence parameters are close to those for perfect acceptance. 
The additional systematic uncertainties on $\kappa^{K3\pi}$ and $\delta_D^{K3\pi}$, when interpreting the four-body results reported here, are $\pm\,0.01$ and $\pm\,2.3^\circ$. 
In a similar study, the additional systematic uncertainty associated with the modulation of the \CP fraction $F_+^{4\pi}$ by the LHCb acceptance is estimated to be $\pm\,0.02$.

It has been shown that \D-mixing effects must be taken into account when using these \CP observables in the determination of \g~\cite{Rama:2013voa}.
The correction is most important in the ADS observables of $\Bm \to D \pim$ decays and is corrected for using knowledge of the decay-time acceptance. From simulation samples, a decay-time acceptance function is defined for both the two-body and four-body \D-mode selections. 
The \D-mixing coefficient $\alpha$, defined in~\cite{Rama:2013voa}, is found to be $-0.59$ and $-0.57$ for the two- and four-body cases, with negligible uncertainties compared to those of the $x$ and $y$ \D-mixing parameters .

\section{Discussion and conclusions}
\label{sec:conclusions}

World-best measurements of \CP observables in $\Bm\to Dh^-$ decays are obtained with the \D meson reconstructed in $\Km\pip$, $\Kp\Km$, $\pip\pim$, $\pim \Kp$, $\Km\pip \pip \pim$ and $\pim \Kp \pip \pim$ final states; this supersedes earlier work~\cite{Aaij:1432548,Aaij:1529913}. 
Measurements exploiting the four-pion \D decay are reported for the first time with the $\Bm \to [\pip\pim\pip\pim]_D\Km$ decay showing an indication of \CP violation at the $2.7\,\sigma$ level.
The charge asymmetry in this mode is positive, similar to the classic GLW modes, $\Bm \to [\Kp\Km]_D\Km$ and $\Bm \to [\pip\pim]_D\Km$, in line with expectation for a multi-body \D mode with a \CP fraction greater than 0.5~\cite{Malde:2015mha}.

\begin{figure}[p]
\centering
\includegraphics[width=0.47\textwidth]{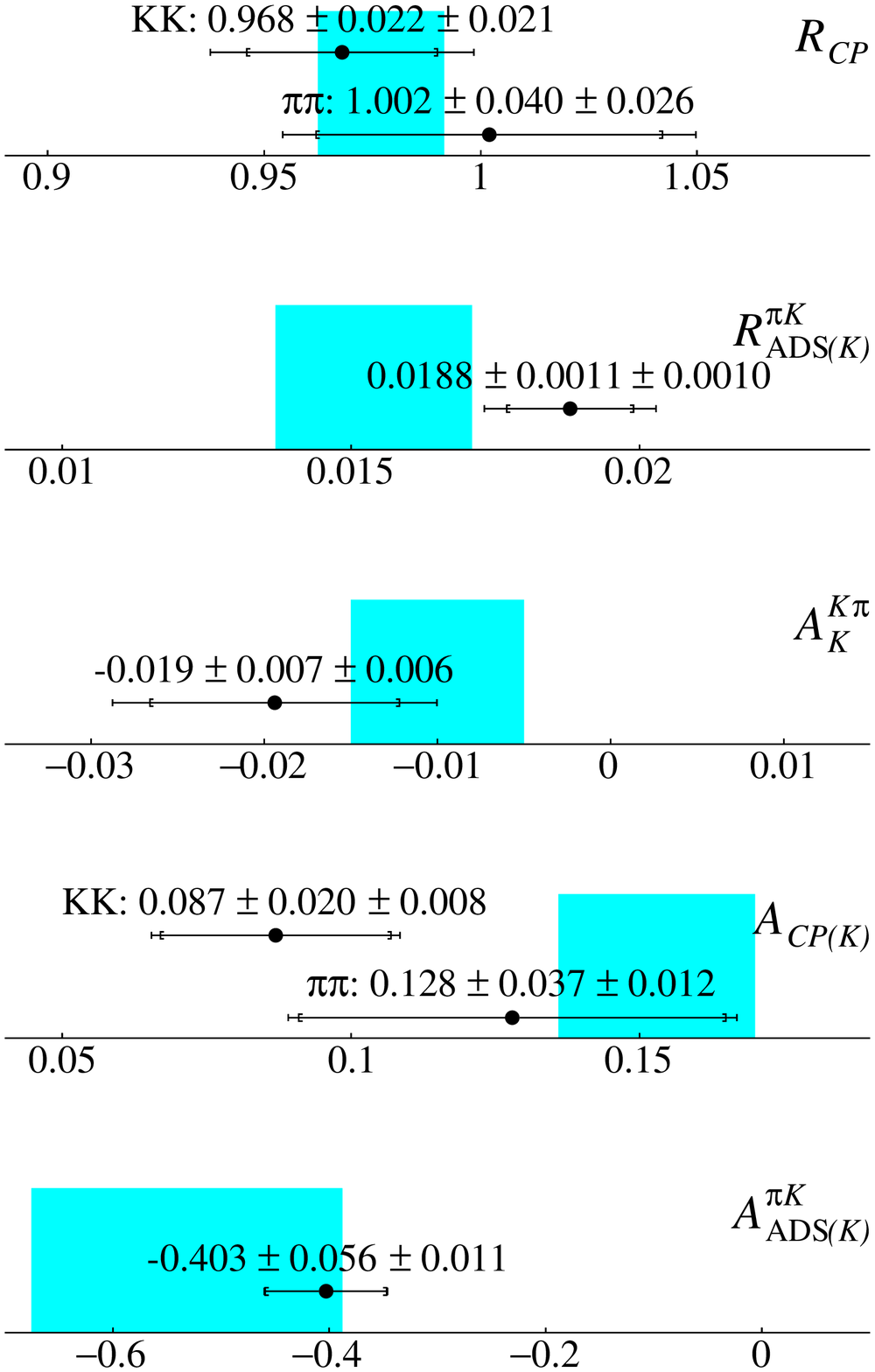}
\includegraphics[width=0.47\textwidth]{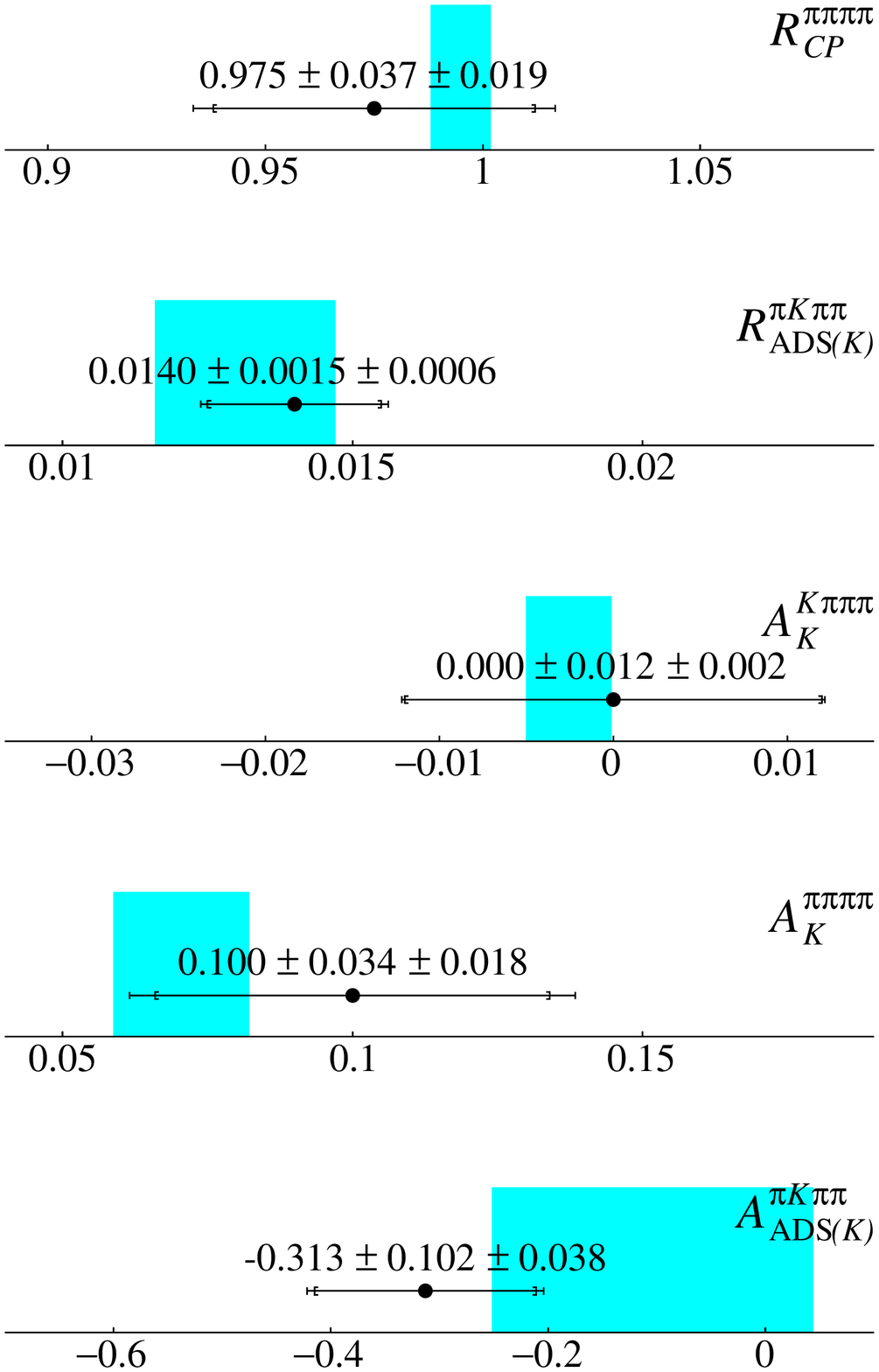} \\
\vspace{1cm}
\includegraphics[width=0.47\textwidth]{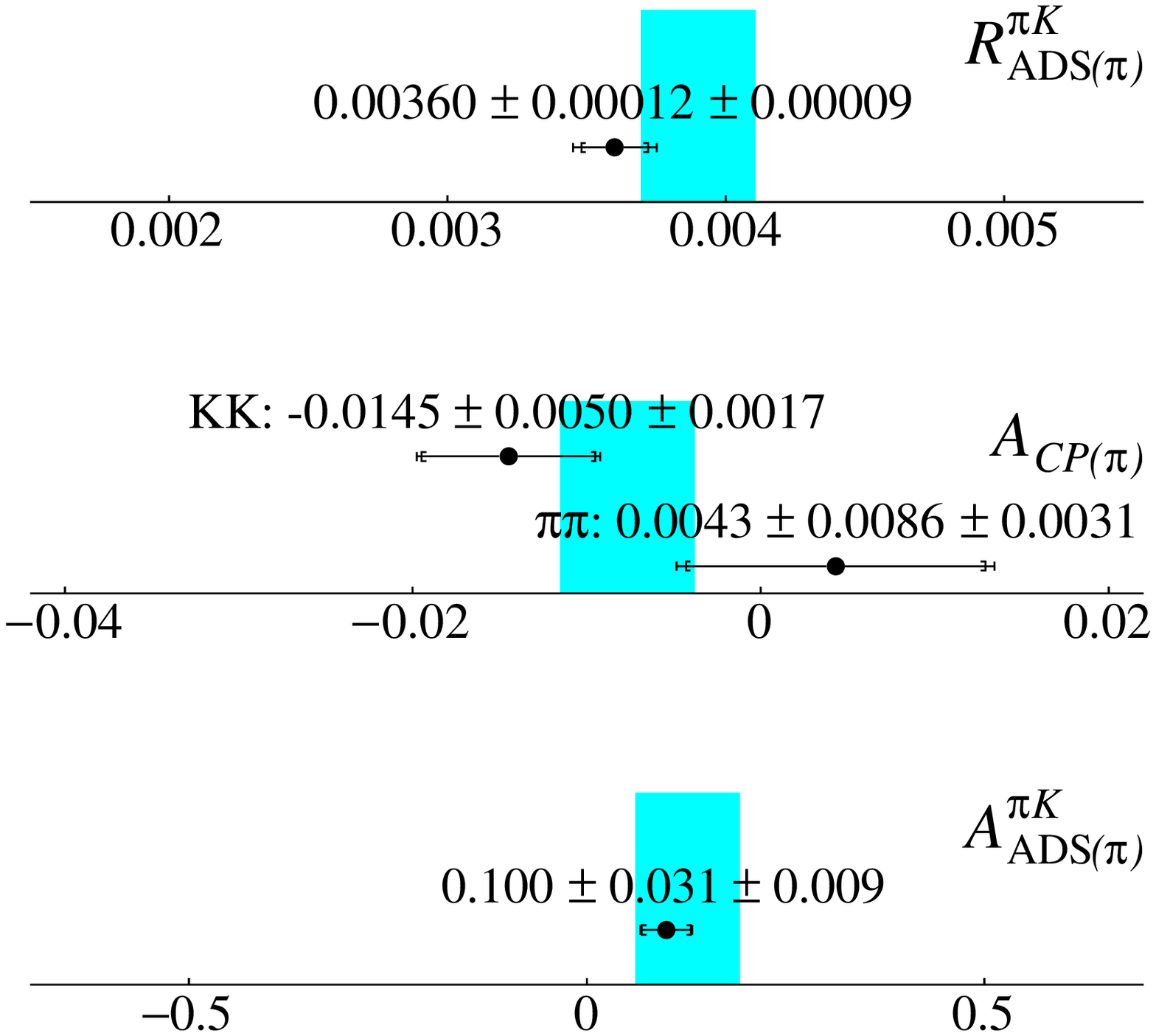}
\includegraphics[width=0.47\textwidth]{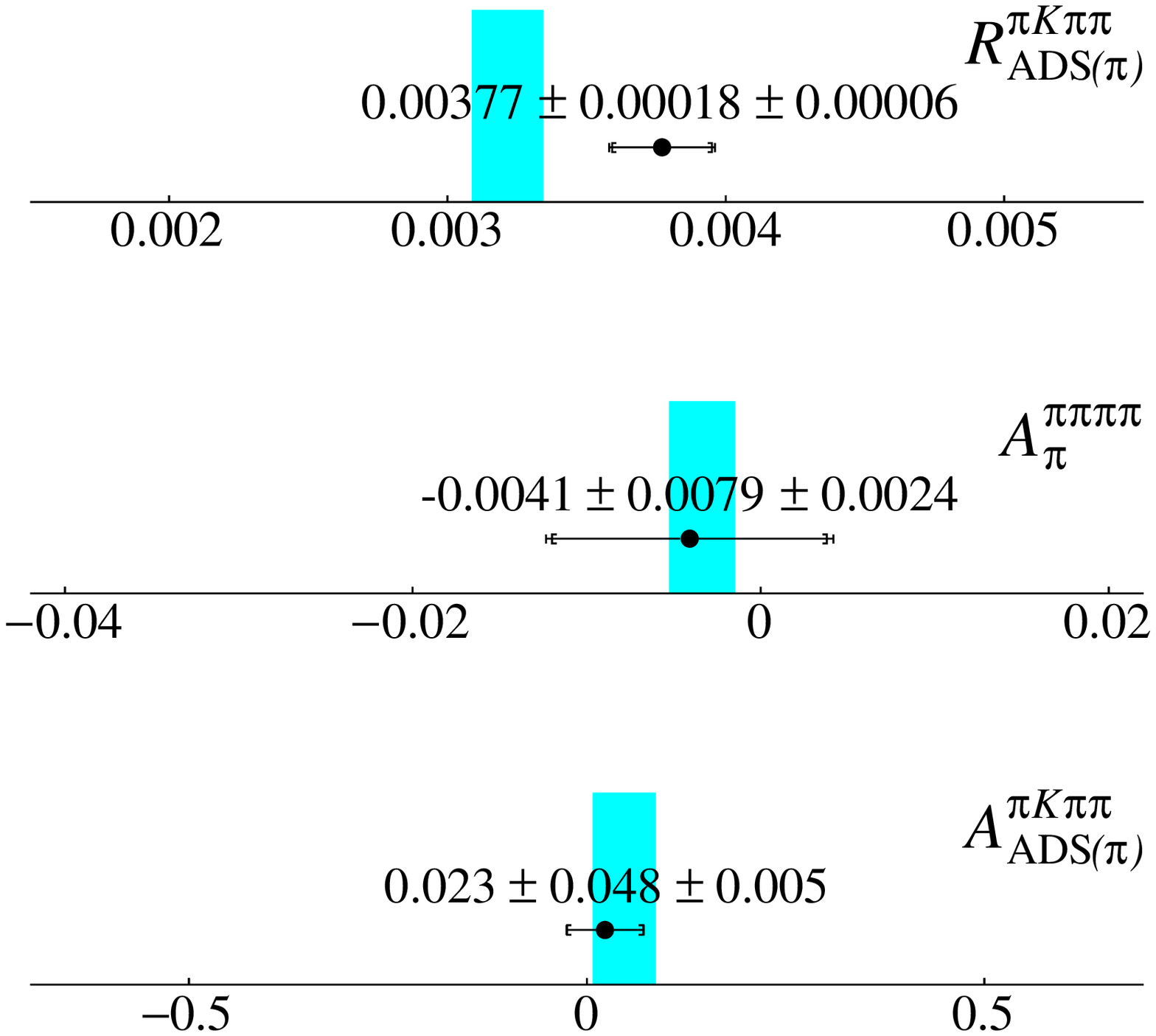}
\caption{Comparison of selected results with the SM expectation (shaded) based on existing knowledge of the underlying parameters and \D-decay measurements, as described in the text. The first five rows show $\Bm\to D\Km$ observables, two-body $D$ decay on the left and four-body on the right. The last three rows show $\Bm\to D\pim$ observables.\label{fig:comparison}}
\end{figure}

A comparison with the SM expectation is made by calculating the \CP observables from current best-fit values of $\g=(73.2^{+6.3}_{-7.0})^\circ$ as well as $\delta_B=(125.4^{+7.0}_{-7.8})^\circ$ and $r_B=(9.70^{+0.62}_{-0.63})\%$ for $\Bpm\to D\Kpm$ decays~\cite{Charles:2004jd}.
For $\Bpm\to D\pipm$ decays, where no independent information on $r_B$ and $\delta_B$ is available, uniform PDFs are used, $180^\circ<\delta_B<360^\circ$ and $0.004< r_B<0.008$.
The \D-decay parameters are taken from the literature: 
$r_D^2=(0.349\pm0.004)\%$ and $\delta_D=(191.8^{+\phantom{1}9.5}_{-14.7})^\circ$~\cite{Amhis:2014hma}; 
$F_+^{4\pi}=0.737\pm0.028$~\cite{Malde:2015mha}; 
$r_D^{K3\pi}=(5.52\pm0.007)\%$, $\delta_D^{K3\pi}=(170^{+37}_{-39})^\circ$ and $\kappa_D^{K3\pi}=0.32^{+0.12}_{-0.08}$~\cite{Evans:2016tlp}. 
The current world averages of the \D-mixing parameters are $x=(0.37\pm0.16)\%$ and $y=(0.66^{+0.07}_{-0.10})\%$~\cite{Amhis:2014hma}, and the $\alpha$ coefficients reported in Sec.~\ref{sec:acc} are used for the small \D-mixing correction. 
For these inputs, the central $68\%$ confidence-level expectation interval is displayed in Fig.~\ref{fig:comparison}, together with the results presented herein. 
It is seen that the \mbox{$\Bm\to D\Km$} measurements are compatible with the SM expectation and that the improvement in the knowledge of the $A_{\text{ADS}}$ observables is particularly significant. 
The measurements presented in this paper improve many of the $C\!P$ observables used in global fits for the unitarity triangle angle $\gamma$ as well as the hadronic parameters $r_B$ and $\delta_B$ for these decays. An improvement in the global best-fit precision on $\gamma$ of around 15\% is anticipated from this work.

\FloatBarrier

\section*{Acknowledgements}
\noindent We express our gratitude to our colleagues in the CERN
accelerator departments for the excellent performance of the LHC. We
thank the technical and administrative staff at the LHCb
institutes. We acknowledge support from CERN and from the national
agencies: CAPES, CNPq, FAPERJ and FINEP (Brazil); NSFC (China);
CNRS/IN2P3 (France); BMBF, DFG and MPG (Germany); INFN (Italy); 
FOM and NWO (The Netherlands); MNiSW and NCN (Poland); MEN/IFA (Romania); 
MinES and FANO (Russia); MinECo (Spain); SNSF and SER (Switzerland); 
NASU (Ukraine); STFC (United Kingdom); NSF (USA).
We acknowledge the computing resources that are provided by CERN, IN2P3 (France), KIT and DESY (Germany), INFN (Italy), SURF (The Netherlands), PIC (Spain), GridPP (United Kingdom), RRCKI and Yandex LLC (Russia), CSCS (Switzerland), IFIN-HH (Romania), CBPF (Brazil), PL-GRID (Poland) and OSC (USA). We are indebted to the communities behind the multiple open 
source software packages on which we depend.
Individual groups or members have received support from AvH Foundation (Germany),
EPLANET, Marie Sk\l{}odowska-Curie Actions and ERC (European Union), 
Conseil G\'{e}n\'{e}ral de Haute-Savoie, Labex ENIGMASS and OCEVU, 
R\'{e}gion Auvergne (France), RFBR and Yandex LLC (Russia), GVA, XuntaGal and GENCAT (Spain), Herchel Smith Fund, The Royal Society, Royal Commission for the Exhibition of 1851 and the Leverhulme Trust (United Kingdom).

\newpage
\appendix
\section*{Appendix: Correlation matrices}
\label{sec:app}

The statistical uncertainty correlation matrices are given in Table~\ref{tab:stat_correlations_kpi_summary} and~\ref{tab:stat_correlations_k3pi_summary} for the 2-body and 4-body fits to data. The correlations between systematic uncertainties are provided in Tables~\ref{tab:syst_correlations_kpi_summary} and \ref{tab:syst_correlations_k3pi_summary}.

\begin{table}[ht]
\caption{Statistical correlation matrix from for the 2-body fit.
\label{tab:stat_correlations_kpi_summary}}
\begin{center}
\scriptsize
\begin{tabular}{l | c c c c c c c c c c c c }
& $A_{K}^{K\pi}$ & $R_{K/\pi}^{K\pi}$ & $A_{K}^{KK}$ & $A_{\pi}^{KK}$ & $R^{KK}$ & $A_{K}^{\pi\pi}$ & $A_{\pi}^{\pi\pi}$ & $R^{\pi\pi}$ & $R_{\text{ADS}(\pi)}^{\pi K}$ & $R_{\text{ADS}(K)}^{\pi K}$ & $A_{\text{ADS}(\pi)}^{\pi K}$ & $A_{\text{ADS}(K)}^{\pi K}$ \\
\hline
$A_{K}^{K\pi}$  & 1  & \phantom{-}0.00  & \phantom{-}0.02  & \phantom{-}0.09  & \phantom{-}0.00  & \phantom{-}0.01  & \phantom{-}0.05  & \phantom{-}0.00  & \phantom{-}0.00  & \phantom{-}0.00  & \phantom{-}0.01  & \phantom{-}0.01 \\
$R_{K/\pi}^{K\pi}$  &     & 1  & \phantom{-}0.00  & \phantom{-}0.00  & -0.32  & \phantom{-}0.00  & \phantom{-}0.00  & -0.18  & \phantom{-}0.01  & -0.11  & \phantom{-}0.00  & \phantom{-}0.00 \\
$A_{K}^{KK}$  &        &             & 1  & -0.01  & -0.01  & \phantom{-}0.00  & \phantom{-}0.02  & \phantom{-}0.00  & \phantom{-}0.00  & \phantom{-}0.00  & \phantom{-}0.00  & \phantom{-}0.00 \\
$A_{\pi}^{KK}$  &        &          &           & 1  & \phantom{-}0.00  & \phantom{-}0.02  & \phantom{-}0.06  & \phantom{-}0.00  & \phantom{-}0.00  & \phantom{-}0.00  & \phantom{-}0.02  & \phantom{-}0.01 \\
$R^{KK}$          &        &           &          &          & 1  & \phantom{-}0.00  & \phantom{-}0.00  & \phantom{-}0.06  & -0.01  & \phantom{-}0.04  & \phantom{-}0.00  & \phantom{-}0.00 \\
$A_{K}^{\pi\pi}$  &        &          &          &          &         & 1  & -0.04  & -0.04  & \phantom{-}0.00  & \phantom{-}0.00  & \phantom{-}0.00  & \phantom{-}0.00 \\
$A_{\pi}^{\pi\pi}$  &        &         &          &          &         &          & 1  & \phantom{-}0.00  & \phantom{-}0.00  & \phantom{-}0.00  & \phantom{-}0.01  & \phantom{-}0.01 \\
$R^{\pi\pi}$           &        &         &          &          &         &          &         &  1 & \phantom{-}0.00  & \phantom{-}0.02  & \phantom{-}0.00  & \phantom{-}0.00 \\
$R_{\text{ADS}(\pi)}^{\pi K}$  &      &      &      &       &       &       &        &          & 1  & -0.02  & -0.04  & \phantom{-}0.00 \\
$R_{\text{ADS}(K)}^{\pi K}$  &      &      &      &       &       &       &        &           &           & 1  & \phantom{-}0.02  & \phantom{-}0.10 \\
$A_{\text{ADS}(\pi)}^{\pi K}$ &      &      &      &       &       &       &        &           &           &          & 1  & -0.05 \\
$A_{\text{ADS}(K)}^{\pi K}$  &      &      &      &       &       &       &        &           &           &          &           & 1 \\
\end{tabular}
\end{center}
\end{table}

\begin{table}[ht]
\caption{Statistical correlation matrix from for the 4-body fit.
\label{tab:stat_correlations_k3pi_summary}}
\begin{center}
\scriptsize
\begin{tabular}{l | c c c c c c c c c c }
& $R_{K/\pi}^{K\pi\pi\pi}$ & $A_{K}^{K\pi\pi\pi}$ & $R_{CP}^{\pi\pi\pi\pi}$ & $A_{\pi}^{\pi\pi\pi\pi}$ & $A_{K}^{\pi\pi\pi\pi}$ & $R_{ADS(K)}^{\pi K \pi\pi}$ & $R_{ADS(\pi)}^{\pi K \pi\pi}$ & $A_{ADS(K)}^{\pi K \pi\pi}$ & $A_{ADS(\pi)}^{\pi K \pi\pi}$ \\
\hline
$R_{K/\pi}^{K\pi\pi\pi}$  & 1 & \phantom{-}0.00  & -0.31  & 0.00  & \phantom{-}0.00  & -0.10  & \phantom{-}0.01  & \phantom{-}0.00  & \phantom{-}0.00 \\
$A_{K}^{K\pi\pi\pi}$  &  & 1   & \phantom{-}0.00  & \phantom{-}0.10  & \phantom{-}0.02  & \phantom{-}0.00  & \phantom{-}0.00  & \phantom{-}0.01  & \phantom{-}0.02 \\
$R_{CP}^{\pi\pi\pi\pi}$    & &  &  1   & -0.00  & -0.02  & 0.04  & \phantom{-}0.00  & \phantom{-}0.00  & \phantom{-}0.00 \\
$A_{\pi}^{\pi\pi\pi\pi}$    & & &   & 1   & -0.02  & \phantom{-}0.00  & \phantom{-}0.00  & 0.01  & 0.02 \\
$A_{K}^{\pi\pi\pi\pi}$    & & & &  & 1   & \phantom{-}0.00  & \phantom{-}0.00  & \phantom{-}0.00  & \phantom{-}0.00 \\
$R_{ADS(K)}^{\pi K \pi\pi}$     & &  &  &  &  & 1   & -0.05  & \phantom{-}0.08  & \phantom{-}0.01 \\
$R_{ADS(\pi)}^{\pi K \pi\pi}$   & & &  &  &  &  & 1    & \phantom{-}0.00  & -0.02 \\
$A_{ADS(K)}^{\pi K \pi\pi}$  &  & & &  &  &  &  & 1  & -0.06 \\
$A_{ADS(\pi)}^{\pi K \pi\pi}$  &  &  &  &  &   &  &  &   & 1 \\
\end{tabular}

\end{center}
\end{table}

\begin{table}
\caption{Correlation matrix for the systematic uncertainties in the 2-body analysis.
\label{tab:syst_correlations_kpi_summary}}
\begin{center}
\scriptsize

\begin{tabular}{l| c c c c c c c c c c c c }
& $A_{K}^{K\pi}$ & $R_{K/\pi}^{K\pi}$ & $A_{K}^{KK}$ & $A_{\pi}^{KK}$ & $R^{KK}$ & $A_{K}^{\pi\pi}$ & $A_{\pi}^{\pi\pi}$ & $R^{\pi\pi}$ & $R_{\text{ADS}(\pi)}$ & $R_{\text{ADS}(K)}$ & $A_{\text{ADS}(\pi)}$ & $A_{\text{ADS}(K)}$ \\
\hline
$A_{K}^{K\pi}$     & 1  & -0.03  & \phantom{-}0.20  & \phantom{-}0.21  & \phantom{-}0.00  & \phantom{-}0.01  & \phantom{-}0.36  & -0.00  & \phantom{-}0.07  & -0.09  & -0.52  & \phantom{-}0.09 \\
$R_{K/\pi}^{K\pi}$&& 1  & \phantom{-}0.09  & -0.10  & -0.10  & -0.06  & -0.03  & \phantom{-}0.24  & -0.01  & -0.22  & -0.03  & -0.13 \\
$A_{K}^{KK}$       &&& 1  & -0.46  & -0.29  & \phantom{-}0.03  & \phantom{-}0.24  & \phantom{-}0.01  & -0.06  & -0.10  & \phantom{-}0.05  & -0.04 \\
$A_{\pi}^{KK}$      &&&& 1  & \phantom{-}0.18  & \phantom{-}0.32  & \phantom{-}0.50  & -0.11  & -0.15  & \phantom{-}0.31  & \phantom{-}0.42  & \phantom{-}0.33 \\
$R^{KK}$              &&&&& 1  & \phantom{-}0.12  & -0.05  & \phantom{-}0.19  & \phantom{-}0.12  & \phantom{-}0.32  & \phantom{-}0.07  & \phantom{-}0.23 \\
$A_{K}^{\pi\pi}$     &&&&&  & 1  & \phantom{-}0.05  & -0.30  & -0.33  & \phantom{-}0.39  & \phantom{-}0.38  & \phantom{-}0.30 \\
$A_{\pi}^{\pi\pi}$    &&&&&&  & 1  & -0.03  & \phantom{-}0.07  & -0.06  & \phantom{-}0.25  & \phantom{-}0.11 \\
$R^{\pi\pi}$            &&&&&&&  & 1  & \phantom{-}0.18  & -0.16  & -0.11  & -0.06 \\
$R_{\text{ADS}(\pi)}$  &&&&&&&&  & 1  & -0.57  & -0.56  & -0.44 \\
$R_{\text{ADS}(K)}$    &&&&&&&&&  & 1  & \phantom{-}0.56  & \phantom{-}0.76 \\
$A_{\text{ADS}(\pi)}$   &&&&&&&&&&  & 1  & \phantom{-}0.40 \\
$A_{\text{ADS}(K)}$     &&&&&&&&&&&  & 1 \\
\end{tabular}

\end{center}
\end{table}

\begin{table}
\caption{Correlation matrix for the systematic uncertainties in the 4-body analysis.
\label{tab:syst_correlations_k3pi_summary}}
\begin{center}
\scriptsize
\begin{tabular}{l| c c c c c c c c c c }
& $R_{K/\pi}^{K\pi\pi\pi}$ & $R_{CP}^{\pi\pi\pi\pi}$ & $R_{ADS(K)}^{\pi K\pi\pi}$ & $R_{ADS(\pi)}^{\pi K\pi\pi}$ & $A_{B_{u}}$ & $A_{K}^{K\pi\pi\pi}$ & $A_{ADS(K)}^{\pi K\pi\pi}$ & $A_{ADS(\pi)}^{\pi K\pi\pi}$ & $A_{DK}^{\pi\pi\pi\pi}$ & $A_{D\pi}^{\pi\pi\pi\pi}$ \\
\hline
$R_{K/\pi}^{K\pi\pi\pi}$  & 1  & \phantom{-}0.11  & -0.04  & \phantom{-}0.13  & -0.01  & \phantom{-}0.00  & -0.13  & \phantom{-}0.17  & -0.14  & \phantom{-}0.08 \\
$R_{CP}^{\pi\pi\pi\pi}$  & & 1  & \phantom{-}0.04  & -0.06  & \phantom{-}0.01  & \phantom{-}0.02  & -0.04  & -0.04  & \phantom{-}0.07  & -0.07 \\
$R_{ADS(K)}^{\pi K\pi\pi}$  && & 1  & \phantom{-}0.14  & -0.00  & \phantom{-}0.02  & \phantom{-}0.87  & \phantom{-}0.10  & \phantom{-}0.03  & \phantom{-}0.01 \\
$R_{ADS(\pi)}^{\pi K\pi\pi}$  &&& & 1  & -0.04  & -0.02  & \phantom{-}0.05  & \phantom{-}0.46  & -0.35  & \phantom{-}0.24 \\
$A_{B_{u}}$  &&&& & 1  & -0.36  & -0.05  & -0.42  & -0.03  & -0.64 \\
$A_{K}^{K\pi\pi\pi}$  &&&&& & 1  & \phantom{-}0.02  & \phantom{-}0.05  & \phantom{-}0.09  & \phantom{-}0.32 \\
$A_{ADS(K)}^{\pi K\pi\pi}$  &&&&&& & 1  & -0.09  & -0.04  & \phantom{-}0.02 \\
$A_{ADS(\pi)}^{\pi K\pi\pi}$  &&&&&&& & 1  & -0.34  & \phantom{-}0.43 \\
$A_{DK}^{\pi\pi\pi\pi}$  &&&&&&&& & 1  & \phantom{-}0.31 \\
$A_{D\pi}^{\pi\pi\pi\pi}$  &&&&&&&&& & 1 \\
\end{tabular}
\end{center}
\end{table}
\clearpage

\newpage
\addcontentsline{toc}{section}{References}
\setboolean{inbibliography}{true}
\bibliographystyle{LHCb}
\bibliography{main}
 
\newpage
\centerline{\large\bf LHCb collaboration}
\begin{flushleft}
\small
R.~Aaij$^{39}$, 
C.~Abell\'{a}n~Beteta$^{41}$, 
B.~Adeva$^{38}$, 
M.~Adinolfi$^{47}$, 
Z.~Ajaltouni$^{5}$, 
S.~Akar$^{6}$, 
J.~Albrecht$^{10}$, 
F.~Alessio$^{39}$, 
M.~Alexander$^{52}$, 
S.~Ali$^{42}$
% ...

{\footnotesize \it
% ...
$ ^{5}$Clermont Universit\'{e}, Universit\'{e} Blaise Pascal, CNRS/IN2P3, LPC, Clermont-Ferrand, France\\
$ ^{6}$CPPM, Aix-Marseille Universit\'{e}, CNRS/IN2P3, Marseille, France\\
% ...
$ ^{10}$Fakult\"{a}t Physik, Technische Universit\"{a}t Dortmund, Dortmund, Germany\\
% ...
$ ^{38}$Universidad de Santiago de Compostela, Santiago de Compostela, Spain\\
$ ^{39}$European Organization for Nuclear Research (CERN), Geneva, Switzerland\\
% ...
$ ^{41}$Physik-Institut, Universit\"{a}t Z\"{u}rich, Z\"{u}rich, Switzerland\\
$ ^{42}$Nikhef National Institute for Subatomic Physics, Amsterdam, The Netherlands\\
% ...
$ ^{47}$H.H. Wills Physics Laboratory, University of Bristol, Bristol, United Kingdom\\
% ...
$ ^{52}$School of Physics and Astronomy, University of Glasgow, Glasgow, United Kingdom\\
% ...
}
\end{flushleft}

\newpage
\end{document}